\documentclass[graybox]{svmult}

\usepackage{type1cm}        
\usepackage{makeidx}         
\usepackage{graphicx}        
\usepackage{multicol}        
\usepackage[bottom]{footmisc}

\usepackage{newtxtext}       %
\usepackage[varvw]{newtxmath}       

\makeindex             

\usepackage{amsmath,amssymb}

\begin{document}

\def\eg{{\em e.g.}}
\def\ie{{\em i.e.}}
\def\wrt{{\em w.r.t.}}
\def\etal{{\em et al.}}
\def\Meff{M_{\rm eff}}
\def\Neq{N_{\cal Q}^{\rm eq}}
\def\qbar{{\bar q}}
\def\Qbar{{\bar Q}}
\def\cbar{{\bar c}}
\def\bbar{{\bar b}}
\def\vrel{v_{\rm rel}}
\def\Lc{{\Lambda_c^+}}
\def\Lb{{\Lambda_b^0}}
\def\Lqcd{{\Lambda_{\rm QCD}}}
\newcommand{\dd}{\mathrm{d}}
\newcommand{\lsim}{\lesssim}
\newcommand{\gsim}{\gtrsim}
\newcommand{\psip}{\psi(2S)}
\newcommand{\jpsi}{J/\psi}
\newcommand{\chic}{\chi_c}
\newcommand{\raa}{R_{\rm AA}}
\newcommand{\npart}{N_{\rm part}}
\newcommand{\pT}{p_{\rm T}}
\newcommand{\pt}{p_{\rm t}}
\newcommand{\Td}{T_{\rm diss}}
\newcommand{\Tpc}{T_{\rm pc}}
\newcommand{\Th}{T_{\rm H}}
\newcommand{\Ds}{{\cal D}_s}
\newcommand{\Ncoll}{N_{\rm coll}}
\newcommand{\beq}{\begin{equation}}
\newcommand{\eeq}{\end{equation}}
\newcommand{\bea}{\begin{eqnarray}}
\newcommand{\eea}{\end{eqnarray}}

\newcommand{\RR}[1]{\textcolor{purple}{#1}}

\title*{Quark Recombination}

\author{Rainer J.\ Fries, Vincenzo Greco and Ralf Rapp}

\institute{Rainer J.\ Fries \at  Cyclotron Institute and Department of Physics and Astronomy, Texas A\&M University, College Station TX 77843, USA;  \email{rjfries@tamu.edu}
\and Vincenzo Greco   \at Department of Physics and Astronomy "Ettore Majorana", University of Catania, and INFN-LNS, Via S. Sofia 64, I-95123 Catania, Italy; \email{greco@lns.infn.it}
\and Ralf Rapp   \at Cyclotron Institute and Department of Physics and Astronomy, Texas A\&M University, College Station TX 77843-3366, USA;   \email{rrapp@tamu.edu}}

%
%
\maketitle

\abstract{Hadronization is a fundamental process occurring at a distance scale of about $1\,\rm fm \simeq \Lqcd^{-1} $, hence within non-perturbative dynamics. In elementary collisions, like $e^+e^-$, $e^-p$, or $pp$, phenomenological approaches to hadronization have been developed based on “vacuum-like” dynamics that require the creation of quark-antiquark and/or diquark pairs during the hadronization process. 
In the 2000s, the idea was developed that in ultra-relativistic nucleus-nucleus (AA) collisions, which lead to the formation of a partonic medium with large (anti-)quark densities, hadronization can occur through the recombination of in-medium quarks, unlike the situation in $e^+e^-$, $e^-p$, and $pp$. 
We give an overview of the main features that characterize quark recombination and have enabled a description of several important experimental observables at both RHIC and LHC over the last two decades. We highlight some additional developments and open issues. We specifically discuss the impact of coalescence on the study of heavy-flavor hadronization, including recent developments showing signatures of (the onset of) quark coalescence even in $pp$ collisions at TeV energies.
Furthermore, we highlight specific features of hadronization for quarkonium in AA collisions, where it has been possible to develop a dynamical kinetic approach that allows to extract more detailed information about the temperature dependence of the heavy-quark interaction in hot QCD matter.}

\section{Introduction and Historical Overview}
\label{sec:1}

 The theory of the strong interaction, Quantum Chromodynamics (QCD), is formulated in terms of the fundamental quark and gluon degrees of freedom (partons), but the observed particles are restricted to hadrons. The latter emerge from the assembly of the 3 (8) color charges of the quarks (gluons) to form color-neutral singlets. Nevertheless, parton degrees of freedom have been established in various experiments. A famous example is electron-positron annihilation into hadrons, where the cross section above invariant masses of $\sim$1.5 GeV is determined by the number of active quark flavors, including their color degeneracy.
 Signatures of partons can also be resolved in hard scatterings of quarks and gluons, \ie,  when momentum transfers are (well) above $\gtrsim$ 1 GeV/$c$ where asymptotic freedom of QCD ensures the interactions to be relatively weak; the original partonic scatterings can then manifest themselves as ``jet" structures in experiment (see \cite{Ali:2010tw} and references therein). Another environment where partonic degrees of freedom are active is Quark-Gluon Plasma (QGP) which can be formed in high-energy collisions of atomic nuclei. Lattice-QCD (lQCD) computations of the finite-temperature partition function (at vanishing baryon chemical potential, $\mu_B$=0) indicate the presence of a crossover transition around a pseudo-critical temperature of 155-160\,MeV associated with the restoration of the spontaneously broken chiral symmetry. However, the liberation of quark and gluon degrees of freedom is likely a rather gradual process whose microscopic description remains a challenging task to date~\cite{Bazavov:2023xzm,Liu:2017qah}.

In both elementary and heavy-ion collision experiments the question arises how hadrons are re-formed from quarks and gluons as dictated by the confinement property of QCD. For the former, in large $Q^2$ processes leading to high-momentum partons, the hadronization essentially occurs in vacuum and is routinely parameterized by ``universal" fragmentation functions. However, at low momentum, the possibility arises that the produced parton finds phase space overlap with surrounding partons to ``recombine" into a hadron. Alternatively, 
hadron formation has been modeled by the formation of ``strings" that fragment into hadrons following energy considerations and quantum number conservation. Finally, more macroscopic models have also been developed, most notably the statistical hadronization model where the hadron production itself is governed by their ``thermal" phase space characterized by an effective temperature, $T_H$. Note that this temperature, at least in elementary collisions, does not represent a thermalization in each event but rather the result of a superposition of many statistically independent events that combine to populate phase space following a ``maximal entropy" principle~\cite{Hagedorn:1984uy,Becattini:2009sc}.  
On the contrary, in (central) heavy-ion collisions ample evidence has arisen that local thermalization happens on an event-by-event basis, \eg, through the success of hydrodynamic modeling~\cite{Heinz:2013th} and measurements of thermal radiation~\cite{NA60:2008ctj}. This allows to a large extent to supersede the microscopic description of the bulk QCD medium by modeling through fluid dynamics  that requires only the knowledge of the equation of state and pertinent transport coefficients to characterize small deviations from equilibrium. In this case, the statistical hadronization model can predict the observed hadron abundances based on chemical equilibration in each event~\cite{Braun-Munzinger:2003pwq}, and its hadronization temperature could be seen as a ``true" temperature in the underlying phase space. However, away from equilibrium, 
a microscopic modeling of hadronization is needed that accounts for the non-universal aspects of the environment and this can be expected to manifest more clearly at higher momenta 
and/or for more exclusive observables beyond the integrated hadron yield.
This includes
the aforementioned models based on strings \cite{Andersson:1983ia,Sjostrand:1993yb,Sjostrand:2006za}
or clusters \cite{Marchesini:1991ch}, 
as well as quark recombination (or coalescence) approaches where hadrons are directly formed from the available quarks and antiquarks \cite{Biro:1994mp,Greco:2003xt,Fries:2003vb,Fries:2008hs}. Note that in the former gluonic configurations are the main building blocks while the latter emphasize quark degrees of freedom (where gluons are often forced to convert into quark-antiquark pairs first).

In this article we provide a brief review of the role that quark recombination models have played in our understanding of hadronization processes in hadronic collisions, highlighting both their successes and drawbacks along with future perspectives on how to overcome the latter. 
We start this review with an overview of the available 
models in Sec.~\ref{sec:2}, including both small collision and large (heavy-ion) 
collision systems for both light- and heavy-flavor (HF) hadron production.
In Sec.~\ref{sec:3} we review the current experimental status of interpretations based on quark recombination. 
In Sec.~\ref{sec:4} we provide a perspective on recombination from the quarkonium sector in an attempt to connect the kinetic transport models that have been developed in this context to the hadronization models discussed in the preceding sections. 
A summary and conclusion are given in Sec.~\ref{sec:5}.

\section{Theoretical Approaches to Quark Recombination}
\label{sec:2}
We first review hadron-production models in elementary-collision systems (with electrons and/or protons), in a somewhat historical perspective,  in Sec.~\ref{sec:1}. We proceed to light-hadron and HF hadron production in heavy-ion collisions in Secs.~\ref{sec:2-2} and \ref{sec:2-3}, respectively. We come back to $pp$ collisions in Sec.~\ref{sec:2-4} to highlight recent developments in the heavy-flavor sector.
\subsection{Small Collision Systems}
\label{sec:2-1}
The fundamental transition in $e^+e^-$, $e^-p$, and $pp$ collisions from the production of a system of partons (quarks and gluons) into hadrons has been studied and modeled by phenomenological approaches already starting from the mid-1970s. The main approach was based on the process of string fragmentation, followed by Monte Carlo event generators at high energy in the following decade, and the development of PYTHIA~\cite{Sjostrand:1993yb,Sjostrand:2006za}, by coupling jet fragmentation with the Lund string fragmentation model. Likewise the cluster hadronization model and
HERWIG~\cite{Marchesini:1991ch} were developed. These approaches were capable of reproducing several features of hadron production data in $e^+ e^-$, $e^-p$ and $pp$ experiments.
The basic idea is that hadronization occurs via the formation of strings and their subsequent fragmentation, where the concept of a ``string" represents the chromo-magnetic field formed between a leading color charge and an anti-color charge. These string, carrying in general a large invariant mass, are assumed to break into hadrons through a quantum tunneling (Schwinger mechanism) of quarks (and diquarks) with a probability given by $\exp\left(-\pi m^2_{\perp q}/{\kappa}\right)$, where $m_{\perp q}$ is the transverse mass of quarks or diquarks and $\kappa \sim 1$~GeV/fm is the string tension. A recent concise review of this approach can be found in Ref.~\cite{Altmann:2024kwx}.
For processes at high $\pt \gg \Lambda_{\text{QCD}}$ (where $\Lambda_{\text{QCD}}\simeq 0.2-0.3$\,GeV represents the basic scale of QCD) it is possible to calculate the momentum distribution $f_a(\pt)$ of a parton $a$ ($=q,\qbar,g$) to be created just after the high $Q^2 \sim \pt^2$ scattering within pQCD. This has been supplemented by the concept of  an ``independent fragmentation function" that effectively describes how the leading parton of momentum $\pt$ converts into hadrons $h$  carrying a fraction $z=\pT/\pt$ of the parton momentum; it can be written as
\cite{Owens:1986mp}
\beq
\label{eq:fragm}
f_H(\pT)= \sum_{a=g,q,\qbar} f_{a} (\pt) \otimes D_{a \rightarrow h}(z)
\eeq
where $D_{a \rightarrow h}(z)$ is the fragmentation function encoding the probability that a hadron of type $h$ is formed from a parton of type $a$. A feature shared by both the Lund fragmentation and by the independent fragmentation approach is that the hadronization process is independent of the environment, \ie, independent of the presence of other strings or quarks and gluons. 

The idea that hadronization of a quark may be affected by the presence of other quarks, and occurs through recombination of these quarks was already present toward the end of the 1970s for the case of $pp$ collisions in the forward rapidity region \cite{Das:1977cp}. It was shown that recombination can describe the longitudinal momentum distribution of pions and kaons in the forward and backward regions. In those regions hadronization appears to keep track of the flavor of the parton distribution function in the protons, an effect known as the leading particle effect. 
This became more evident in heavy-flavor production as an asymmetry in $D^+$/ $D^-$ production in the forward region was seen by WA82 at CERN \cite{WA82:1993ghz} and by E791 at FNAL \cite{E791:1996htn} in $\pi^--$nucleus collisions. 
In fact, the asymmetry $A=\frac{N_{D^-}-N_{D^+}}{N_{D^-}+N_{D^+}}$ can be expected to go to zero in a fragmentation picture while it has been measured to go toward unity in the very forward direction, as discussed further in Sec.\ \ref{sec:3}. 
In the early 2000s, this effect could be quantitatively explained as a recombination of the $\cbar$ produced in the collision with a $d$ valence quark from the $\pi^-$ beam remnants. It is favored \wrt\, the $c + \bar{d} $ recombination, which involves only a sea quark from the $\pi^-$\cite{Braaten:2002yt,Rapp:2003wn}. We notice that this effect is only visible in a quite restricted region of the cross section and is induced by the parton composition of the initial colliding hadrons.

\subsection{Light-Hadron Production in Heavy-Ion Collisions}
\label{sec:2-2}
Early ideas that the quark content of the QGP is imprinted on the hadron abundances in the final-state of heavy-ion collisions were developed within a statistical hadronization model \cite{Letessier:1998sz}. Here, the hadron abundances are governed by the chemical-freezeout temperature, but the chemical potentials (or fugacities) characterizing the conservation laws in the system are evaluated in the partonic phase. A somewhat more microscopic description with more explicit reference to quark coalescence was proposed in terms of the ALCOR (Algebraic Coalescence Rehadronization) model \cite{Biro:1994mp}, where bulk hadronization was computed from the underlying quark content and the resulting hadron ratios were obtained from a schematic rate equation network with an effective cross section. ALCOR was developed to predict hadron yields ($p_{\text T}$-integrated) and ratios and does not compute the momentum spectra of the different hadrons. The significance of the partonic initial state in these models, however, had ultimately to compete with the success of the statistical hadronization model where all hadron ratios are directly computed in the hadronic system and require less parameters~\cite{Braun-Munzinger:2003pwq}, while merely asserting that the interaction rates through the phase transition are large enough to establish hadro-chemical equilibrium. 

In 2002, the idea that hadronization by quark coalescence can be a dominant mechanism at midrapidity in AA collisions was resurrected. After all, with a dense partonic medium providing a significant probability for quarks to be close enough in phase space to form color-neutral objects, this would appear to be quite natural. But the main challenge that these models ultimately addressed was to identify specific features of the coalescence mechanism to be seen in measurable observables. The first suggested fingerprint of quark coalescence has been the enhancement of the proton-to-pion ratio at intermediate $\pT\simeq3-6$ GeV/$c$ compared to $e^+e^-$ or $pp$ collisions \cite{Hwa:2002tu,Greco:2003xt,Fries:2003vb}. The latter were thought to be described by an independent fragmentation mechanism, which is schematically represented by Eq.~(\ref{eq:fragm}). Recombination, on the other hand, can be thought of as a folding of the product of quark and anti-quark distribution functions with a wave function of the hadron. For mesons it can be sketched as 
\beq
\label{eq:coal-m}
f_M(\vec{x},\vec{p})= f_{q} (\vec{x}_a,\vec{p}_a) \otimes f_{\qbar} (\vec x_b,\vec p_b) \otimes W_M (\vec{x}_{ab},\vec{q}_{ab})\delta(\vec{p}-\vec{p}_a-\vec{p}_b)
\eeq
where $f_{q,\qbar}(\vec x,\vec p)$ are the quark (anti-quark) distribution function and  $W_H (\vec x_{ab},\vec q_{ab})$ is the hadron Wigner function depending on the relative quark coordinates. For baryons it reads as:
\bea
\label{eq:coal-b}
f_B(\vec x,\vec p)&=&f_{q} (\vec x_a, \vec p_a) \otimes f_{q} (\vec x_b, \vec p_b) \otimes f_{q} (\vec x_c,\vec p_c)\otimes W_B (\vec x_{ab},\vec x_{bc}, \vec q_{ab}, \vec q_{bc})\nonumber\\
&\times& \, \delta(\vec{p}-\vec{p}_a-\vec{p}_b-\vec{p}_c)
\eea
where the folding includes the integration over the phase space and a statistical factor counting statistical probability that quark colors match to form a color-neutral object with the spin and isospin of the considered hadron~\cite{Greco:2003mm,Fries:2003kq,Kordell:2021prk}.

In a coalescence approach baryon production can be seen as favored \wrt \, independent fragmentation because it is not required that a diquark is created from the vacuum, but rather the combination of close-by quarks is exploited. More specifically, for the case of ultra-relativistic heavy-ion collisions, a thermal parton spectrum is expected at low $\pt$, and this would make the production of baryons similar to that of mesons because the product of an exponential is equal to the exponential of the sum $f_M(P)\sim f_B(P) \sim e^{-\sum_i p_i/T}$ where $P=\sum_i p_i$, is the sum of two parton momenta for mesons and three for baryons. Of course in a realistic scenario one has to consider the spin, isospin, and color degeneracy factors for the specific baryon and, especially for lower-$p_\text{T}$ mesons, the effect of finite masses, hadron wave functions, radial flow and feed-down from resonance decays.
One can qualitatively obtain that, even if recombining three quarks should have a smaller probability than recombining two quarks, for a baryon of momentum $P$ one has to draw on a parton distribution at about $P/3$ where the density is exponentially larger with respect to the larger $P/2$ leading to a meson of the same momentum $P$~\cite{Fries:2008hs}. Furthermore, the total probability of recombination to form a hadron at each momentum $P$ is in first approximation independent on the shape of the wave function that fully factorizes out for a massless thermal distribution. One should also  notice that for a power law parton spectrum, $f_q (p)\sim p^{-\beta}$, as occurs for partons at high $\pt$, the coalescence mechanism is not very efficient because parametrically $f_H(P)\sim (P/n_q)^{-\beta n_q}$ generates a very steep spectrum especially for baryons. Instead, the independent fragmentation would only result in a shift toward lower momenta of the power law spectrum, $f_H(P)\sim P^{- \beta - \delta}$, with $\delta << \beta$.

The general scheme of hadronization implies an important pattern between the elliptic flows, $v_2= \langle \cos(2\,\phi_p)\rangle$, of mesons and baryons. Under the approximation of the same collinear momentum (more generally: velocity) of quarks, Eqs.~(\ref{eq:coal-m})
and (\ref{eq:coal-b}) show an elliptic flow angular modulation, $f_q (\vec p_\text{t})\sim f_q(p_\text{t})[1 + 2 \, v_{2q}(p_\text{t}) \,\cos(2\,\phi_p)]$ at the the quark level is transferred to the hadronic level according to:
\bea
\label{eq:vn-scaling}
f_M(\vec p_\text{T}) \sim f_q(\vec p_\text{T}/2) f_{\qbar}(\vec p_\text{T}/2) \sim [1+ 2\, v_{2q}(p_\text{T}/2) \cos(2\,\phi_p)]^2 \nonumber\\
\sim 1+ 4 \, v_{2q}(p_\text{T}/2) \cos(2\,\phi_p)\nonumber\\
f_B(\vec p_\text{T}) \sim f_q(\vec p_\text{T}/3) f_{q}(\vec p_\text{T}/3) f_{q}(\vec p_\text{T}/3) \sim [1+ 2\, v_{2q}(p_\text{T}/3) \cos(2\,\phi_p)]^3 \nonumber\\
\sim 1+ 6 \, v_{2q}(p_\text{T}/3) \cos(2\,\phi_p)
\eea
which can be summarized in terms of a quark number scaling of the elliptic flow, \ie,  a collapse into a universal curve of
$1/n_q \,v_{2H}(p_\text{T}/n_q)$ for all hadron species $H$~\cite{Molnar:2003ff}. This feature of hadronization in coalescence was soon confirmed by more realistic approaches that where initially developed to evaluate the baryon and meson spectra at RHIC and in particular the $B$/$M$ ratio at intermediate $p_\text{T}$~\cite{Greco:2003xt,Fries:2003vb}, which has also been confirmed at LHC energy~\cite{Minissale:2015zwa}.
The quark number scaling as sketched here depends on several assumptions (some of which are not realistic): collinear coalescence, vanishing relative velocity, $\vrel$ (or relative momentum for quarks of equal mass) or, equivalently, Dirac-$\delta$ Wigner functions, $\sim \delta(p_q-P_H/n_q)$, no feed-down from resonance decays, and absence of space-momentum correlations in the quark distribution functions. However, already early approaches have shown that an approximate quark number scaling is largely preserved when the above assumptions are relaxed~\cite{Greco:2003mm,Greco:2004ex,Minissale:2015zwa}. The aspect that has been explored less and only in the heavy-quark sector  is the impact of space-momentum correlations that can significantly affect the quark number scaling as well as the $\pT$ range in which coalescence is dominant, especially for baryons~\cite{He:2019vgs}; we will elaborate and discuss this in more in detail in Sec.~\ref{sec:3.2.1} for light flavor and \ref{sec:3.2.2} for heavy flavor.

The main fingerprints of hadronization by coalescence manifest themselves in a large $B/M$ ratio, a higher $\raa$ for baryons \wrt\, to mesons and an approximate scaling of the $v_{2H}(\pT)$, all at intermediate $\pT\sim 2-6$\,GeV/$c$. 
During the last two decades there have been developments of several approaches that may differ, \eg, by the specific form of the hadron Wigner function, the treatment of resonance production and decay and the determination of the quark distribution functions. However,
Eqs.~(\ref{eq:coal-m}) and 
 (\ref{eq:coal-b}) represent the general scheme of hadronization by phase-space coalescence in the Wigner formalism, and the developments toward a more realistic approach have kept the main features discussed above~\cite{Greco:2007nu,Song:2015ykw,Han:2016uhh,Kordell:2021prk,Altmann:2024kwx}. 
A main criticism of coalescence that remains since the beginning has been the violation of energy conservation, because a process of two (or three) on-shell particles cannot conserve energy and momentum when recombining into a bound-state on-shell hadron. A first general justification for a simple phase-space coalescence is that the process does not occur in the vacuum and the medium would participate in the process allowing for energy-momentum conservation, considering also that a constituent-quark picture does not lead to a strong violation of energy conservation for most of the particles considered, except for pions. A fully dynamical microscopic description is certainly missing and not easy to achieve; however, the issue can be expected to be more relevant in the determination of spectra at low $\pT$ and absolute hadronic yields in the light sector. As for pions, whose direct production in a constituent-quark picture implies quite a large violation of energy conservation, it has to be said that realistic approaches consider also the formation of resonances like $\rho$, $\omega$, and $\Delta$ and it has been shown that in such an approach the direct production of pions is not dominant~\cite{Greco:2003xt,Minissale:2015zwa}.

To address the issue of 4-momentum conservation, and the closely related issue of the thermal equilibrium limit, an approach
known as Resonance Recombination Model (RRM) has been developed~\cite{Ravagli:2007xx,Ravagli:2008rt}. Derived from an underlying Boltzmann equation, hadronization is implemented as a scattering process of two quarks into a hadron Breit-Wigner meson resonance with a finite width $\Gamma$ (typically of order of a few hundred MeV). The formulation in terms of the invariant mass-squared, $s=(p_a+p_b)^2$, of the incoming quark and antiquark, enforces 
energy conservation while its kinematic reach is still limited to states close to or above the nominal quark-antiquark threshold (and thus does not apply to, \eg, direct-pion production). The Boltzmann equation can then be evaluated in the steady state and off-equilibrium quark distributions can be substituted into the expression which may be sketched as:
\beq
\label{eq:RRM}
f_M(\vec P) = \int d^3p_a d^3p_b f_{q}(\vec x_a,\vec p_a) \otimes f_{\qbar} (\vec x_b,\vec p_b) \otimes 
\sigma_M (s)\, \vrel(p_a,p_b) \,\delta^{(3)}(\vec P-\vec p_a-\vec p_b) \ .
\eeq 
When  comparing Eqs.~(\ref{eq:RRM}) and (\ref{eq:coal-m}) one recognizes 
that the RRM shares several features with the coalescence approach because the main underlying process is based on the product of quark distribution functions and the sum of quark 3-momenta; the main difference is that the relative momentum and total energy of the quarks determine the formation probability according to a Breit-Wigner
cross section, $\sigma_M(s) \vrel$, and not through a simple overlap with a Wigner transform of the hadron wave function. Nevertheless, the resonant cross section similarly restricts the phase space for hadron formation from recombination through its invariant mass being centered around the resonance mass.
Being based on an underlying Boltzmann equation, a key feature of the RRM approach is that it satisfies the equilibrium limit of the produced hadron when inserting two equilibrium quark distributions. This fully upholds in the presence of a collectively expanding source with radial flow and its angular modulations, in particular the elliptic flow~\cite{He:2010vw}. Consequently, the RRM not only satisfies thermodynamic consistency but also allows for a seamless transition into the hydrodynamic regime at lower $p_\text{T}$~\cite{He:2010vw}. This further implies the possibility for a controlled implementation of space-momentum correlations~\cite{He:2019vgs}, a key hallmark of hydrodynamic simulations for heavy-ion reactions. 
The term $\sigma_M(s) v_{rel}\propto |{\cal M}|^2$ can be written more rigorously as an invariant matrix element squared which eliminates the (spurious) dependence on the relative velocity. This, in turn, enables to systematically go beyond the quasi-particle approximation by accounting for broad spectral functions of the thermal partons, as expected for a strongly interacting medium. In fact, the  presence of resonance interactions prior to recombination naturally connects the physics of a strongly coupled QGP to its hadronization~\cite{Liu:2017qah} as emerging from the same microscopic interaction.
The RRM for mesons, Eq.~(\ref{eq:RRM}), has been extended to baryons considering their formation as two-step processes with an initial doorway state through a recombination into diquarks followed by another quark-diquark recombination~\cite{He:2019vgs}, involving two quarks provided by the surrounding medium (\ie, not from the vacuum). This is also rather natural as diquarks in the color-antitriplet channel emerge from an attractive $qq$ interaction (as does the color-singlet quark-diquark channel).

\subsection{Open Heavy-Flavor Hadronization in Heavy-Ion Collisions}
\label{sec:2-3}
Hadronization via quark recombination has been widely applied to open heavy-flavor (HF) hadrons where a formation by constituent quarks is even better motivated than for light hadrons, especially in the low-$\pT$ region. Already in 2004, it was noted that $D$ mesons would acquire a significant elliptic flow from the recombination with light quarks \cite{Greco:2003vf}. We briefly recall that the elliptic flow measures the anisotropy in the emission of particles with respect to the azimuthal angle $\phi$ and can be defined as $v_2=\langle \cos(2\phi)\rangle$. At that time heavy quarks were thought to be weakly interacting with the QGP medium and hence to have a rather small $v_2(p_\text{T})$ from their diffusion through the QGP (or the suppression in their radiative energy loss). Therefore, the idea was that recombination can set a minimum value for $v_2(p_\text{T})$ of $D$ mesons that at intermediate $\pT$ in semi-central collisions at RHIC was estimated to be as large as $5\%$ under the assumption of a vanishing charm elliptic flow $v_{2c}$. It was also possible to estimate an upper limit of $v_{2D}(\pT) \sim 15-20 \%$ in the "unrealistic case" (according to the standard lore of 20 years ago) of charm quarks flowing with the QGP medium. Soon after, experimental data and theoretical modeling have shown that charm quarks indeed have highly non-perturbative interactions in the hot QCD matter~\cite{vanHees:2004gq} and that the "unrealistic case" was close to the correct one \cite{vanHees:2005wb,vanHees:2007me}; this will be further discussed in Sec.~\ref{sec:3.2.2}.
A main feature of hadronization of a charm quark by recombination with a light quark is to enhance the $v_2$ while reducing the suppression in the $\raa$ of the $D$ meson  \wrt\, the charm quark; this is different from the effect of in-medium energy loss followed by hadronization by fragmentation~\cite{vanHees:2005wb,Scardina:2017ipo}. This additional fingerprint of hadronization by coalescence has been recognized by  various group studying open HF production as quite relevant to achieve an agreement with experimental data~\cite{Rapp:2018qla,Zhao:2023nrz}.
Beyond the impact on the $\pT$ spectrum, \ie, the $\raa(\pT)$ and $v_2(\pT)$ of $D$ mesons, there has been another very relevant aspect of hadronization by coalescence, which is the large charm-baryon production and an enhanced production of charm mesons containing strange (or even bottom) quarks. In $e^+e^-$ and $e^-p$ collisions the charm hadron production has been seen to be dominated by $D$ production, with the 
fraction of charm going to $\Lc$ being quite small, $f(c\rightarrow\Lc)\simeq 0.06$. Therefore, charm-baryon production and the study of charm interaction in the QGP medium (and even in $pp$ collisions) were not priorities until a few years ago.
In 2009, a phase-space coalescence approach to $\Lc$ production predicted a large ratio of $\Lc/D^0 \sim 1$ in AA collisions at RHIC energy~\cite{Oh:2009zj}. Subsequently, other coalescence plus fragmentation approaches confirmed a similar finding at RHIC and also anticipated a moderate decrease of the ratio at intermediate $\pT$ at the LHC~\cite{Plumari:2017ntm}; a similar finding was made in the RRM approach~\cite{He:2019vgs}. However, it has to be said that such predictions are obtained by enforcing that in the limit $\pt \rightarrow 0$, all charm quarks hadronize by coalescence~\cite{Oh:2009zj,Plumari:2017ntm,Cao:2019iqs,Altmann:2024kwx}, which is significantly different from nucleon coalescence for light-nuclei production.
However, these predictions were qualitatively confirmed by the STAR collaboration in 2019~\cite{STAR:2019ank} and more recently by ALICE~\cite{ALICE:2021bib}, see Sec.~\ref{sec:3} for a more detailed discussion. 

Another approach to in-medium hadronization that utilizes the idea of quark recombination has been developed within POWLANG modeling of open HF production~\cite{Beraudo:2022dpz}. Here each (anti-)charm quark can randomly recombine with quarks or diquarks which are distributed according to a thermal spectrum at $T_{\rm H}$=155\, MeV.
The recombination occurs if the invariant
mass of the pair is larger than the mass of the lightest charmed hadron; subsequently, the cluster decays into a charmed
hadron with the same baryon number and strangeness of
the cluster~\cite{Beraudo:2022dpz,Beraudo:2023nlq}. The main difference with the coalescence approach is that the probability of recombination is not determined by a hadron wave function. However, a key aspect is that only when recombination of 
quark (di-quark) pairs close in space and momentum is considered the approach leads to a large $\Lc/D^0$ close to experimental data. This qualitatively points in the same direction as the coalescence approach of locally recombining objects with not too large invariant (relative) momentum.

More generally, it has been pointed out that the peak and the $\pT$-range where $\Lc/D^0$ remains large depend on the strength of the space-momentum correlation between charm quarks and the medium~\cite{He:2019vgs}. This would also affect the quantitative behavior of the ratio
$v_{2}^\Lc/v_2^D$ vs $\pT$. Such a measurement would allow a better insight into the dynamics of charm quarks and achieve a more solid confirmation of hadronization by recombination.

\subsection{Heavy-Flavor Production in $pp$ Collisions}
\label{sec:2-4}
An intriguing recent finding has been that $\Lc/D^0$ ratio in $pp$ and $pPb$ collisions at TeV collision energies reaches values much larger than those in $e^+e^-$ and $e^-p$, of up to $\sim$0.6 around $\pT \simeq 1.5 $ GeV/$c$~\cite{ALICE:2021rzj,ALICE:2022exq}.
The surprising large value of $\Lc$ has challenged the standard modeling of particle production in $pp$ collisions; in particular, the PYHTIA event generator, where the main mechanism is string fragmentation, has been able to account for the particle production in $e^+e^-$, but failed in its standard version to account for the $\Lc/D^0$ ratio by nearly one order magnitude~\cite{Skands:2014pea}. 

An attempt to apply the coalescence plus fragmentation model developed for AA collisions to describe $pp$ collisions in a hydrodynamical approach \cite{Weller:2017tsr}, \ie, assuming a droplet of QGP medium, one finds quite good agreement with the observed $\Lc/D$ ratio vs.~$\pT$, an agreement that extends also to $\Sigma_c^{0,++}/D^0$ and, within still significant systematic errors, to $\Xi_c^0/D^0$, see the left panel of Fig.\ \ref{fig:Sigmac_D0_pp}.

In a different line of approach, it has been proposed that $\Lc$ production can be enhanced within the statistical hadronization model. In the latter charm-hadron states are populated in relative chemical equilibrium, evaluated at a typical ``hadronization temperature" of $T_{\rm H}$=160-170\,MeV~\cite{He:2019tik} while $c\bar c$ content in the system is fixed by its hard production. In this framework the feed-down from excited charm baryons decaying into the ground-state $\Lc$ is sizable, and the inclusion of known states as quoted by the particle data group (PDG)~\cite{ParticleDataGroup:2018ovx} already enhances the $\Lc/D^0$  to about 0.2-0.25~\cite{Andronic:2007zu,He:2019tik}. The main idea in Ref.~\cite{He:2019tik} was to include excited states that go well beyond the PDG listings, as many ``missing" states are predicted by the relativistic quark model~\cite{Ebert:2011kk} and lattice QCD~\cite{Padmanath:2014lvr}. 
Upon adding these states, the resulting $\Lc/D^0$ ratio increases to 0.45-0.57, quite compatible with experiment. The SHM weights for the production yields were combined with the $\pT$-dependence from fixed-order-next-to-leading-log (FONLL) fragmentation functions which entails an approach to the values measured in $e^+e^-$ collisions at high $\pT$. We note that the SHM approach is not necessarily distinct from the coalescence model, as the former can also be thought of as depending on the available phase space of the surrounding event. In particular, the reduction of the $\Lc/D^0$ ratio in low-multiplicity $pp$ collisions has been reasonably well captured by enforcing exact quantum number conservation within the canonical ensemble~\cite{Chen:2020drg}.   

The large $\Lc/D^0$ has supplied further evidence for a recombination mechanism even when it is studied from the point of view of an approach based on string fragmentation. It was found that also in this framework it is necessary to include the possibility of string color-reconnection (PYTHIA-CR) that allows color neutralization between endpoints of different strings close by~\cite{Christiansen:2015yqa}. Such a mechanism sets in when the string density is large and leads to a recombination of local endpoints with the formation of small invariant-mass objects close to the one of charm hadrons. Even though it starts from a mechanism different from coalescence, PYTHIA-CR goes in the direction of dominance of baryon production through recombination of "existing" quarks and not from the breaking of large invariant-mass strings; see \ref{sec:3.1} for a further discussion on this point and a direct comparison to experimental data.
Furthermore, it is interesting to note that in a coalescence approach even the feed-down $\Lc \leftarrow \Sigma_c^{0,+,++}/\Lc$ has been correctly predicted~\cite{Altmann:2024kwx}.  This is a non-trivial finding considering that PYTHIA, even with CR switched on, largely overestimates this feed-down channel. A more thorough analysis has shown that this is associated with a suppression of the diquark state with angular momentum $l=1$ which favors the production of $\Sigma_c$ resonances that largely decay into $\Lc$~\cite{Altmann:2024kwx}. Such a suppression is needed in the description of heavy-baryon production in $e^+e^-$ collisions, but in $pp$ collisions the data suggest that a simple color reconnection according to SU(3) counting is preferred.
In 2024, the success of a coalescence plus fragmentation framework for open heavy flavor has led to the inclusion of phase-space coalescence for heavy quarks in the Wigner formalism in the well-known Monte Carlo event generator EPOS4 for both $pp$ and AA collisions~\cite{Zhao:2023ucp,Zhao:2024ecc}. In particular, these works have adopted the idea of Ref.~\cite{He:2019tik,He:2019vgs} by evaluating the production yields of resonances using the SHM with an augmented set of excited states as predicted by the relativistic quark model.
The possibility to measure experimentally multi-charm baryon production at ALICE3, e.g.\ $\Xi_{cc}, \Omega_{cc},\Omega_{ccc}$, with a system size scan will allow quite thorough insights into hadronization by coalescence 
and into the degree of charm quark thermalization \cite{Minissale:2023dct}.

Interestingly, elliptic flow could be exploited even in elementary collisions: a finite $v_2(\pT)$ of $D$ meson has been observed in $pp$ and $pA$ collisions at the LHC~\cite{CMS:2020qul,ATLAS:2019xqc,CMS:2018loe}. A robust measurement of the $v_2$ of $\Lc$ relative to $D$-mesons could illuminate the role of initial vs final-state effects for this observable, and ultimately be more sensitive to the contribution from hadronization as the portion from a short-lived QGP in small systems will be relatively smaller compared to AA collisions.
Along these lines there has also been a study of bottom-hadron production for $pp$ collisions at $\sqrt{s}=5\,\rm TeV$, suggesting that at midrapidity coalescence should be the dominant mechanism for $B$-mesons up to $\pT \simeq 6$ GeV/$c$  and for $\Lambda_b$ up to $\pT \simeq 12$ GeV/$c$, leading to a ratio of $\Lb/\bar{B^0}= 0.75 \pm 0.2$ at midrapidity \cite{Minissale:2024gxx}.

The SHM approach has also been applied to the bottom sector, in particular with implementing the idea of a largely extended spectrum of exited baryons relative to the PDG listings~\cite{He:2022tod}. Much like in the charm sector, the feed-down from excited states leads to a  $\Lb/\bar{B^0}$ ratio that is enhanced by $\sim$60\% compared to the case when only PDG states are included.

\section{Experimental Signatures of Recombination}
\label{sec:3}
We have organized this section into a part on small collision systems (Sec.~\ref{sec:3.1}) and on heavy-ion collisions (Sec.~\ref{sec:3.2}), which in both cases encompasses light and heavy flavor hadron observables.

\subsection{Elementary Collision Systems}
\label{sec:3.1}
\begin{figure}[t]
    \centering
 \includegraphics[width=0.53\linewidth]{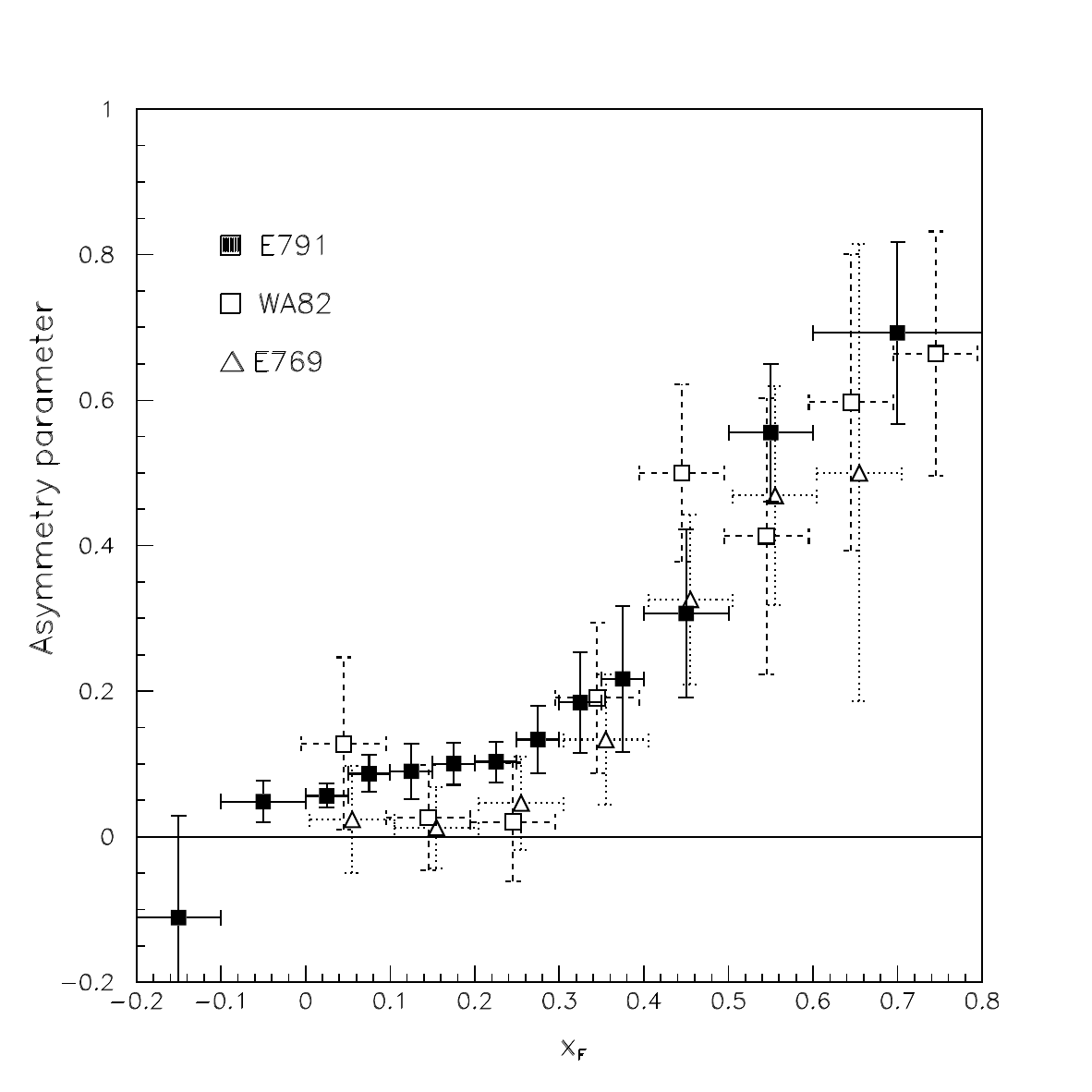}
 \includegraphics[width=0.46\linewidth]{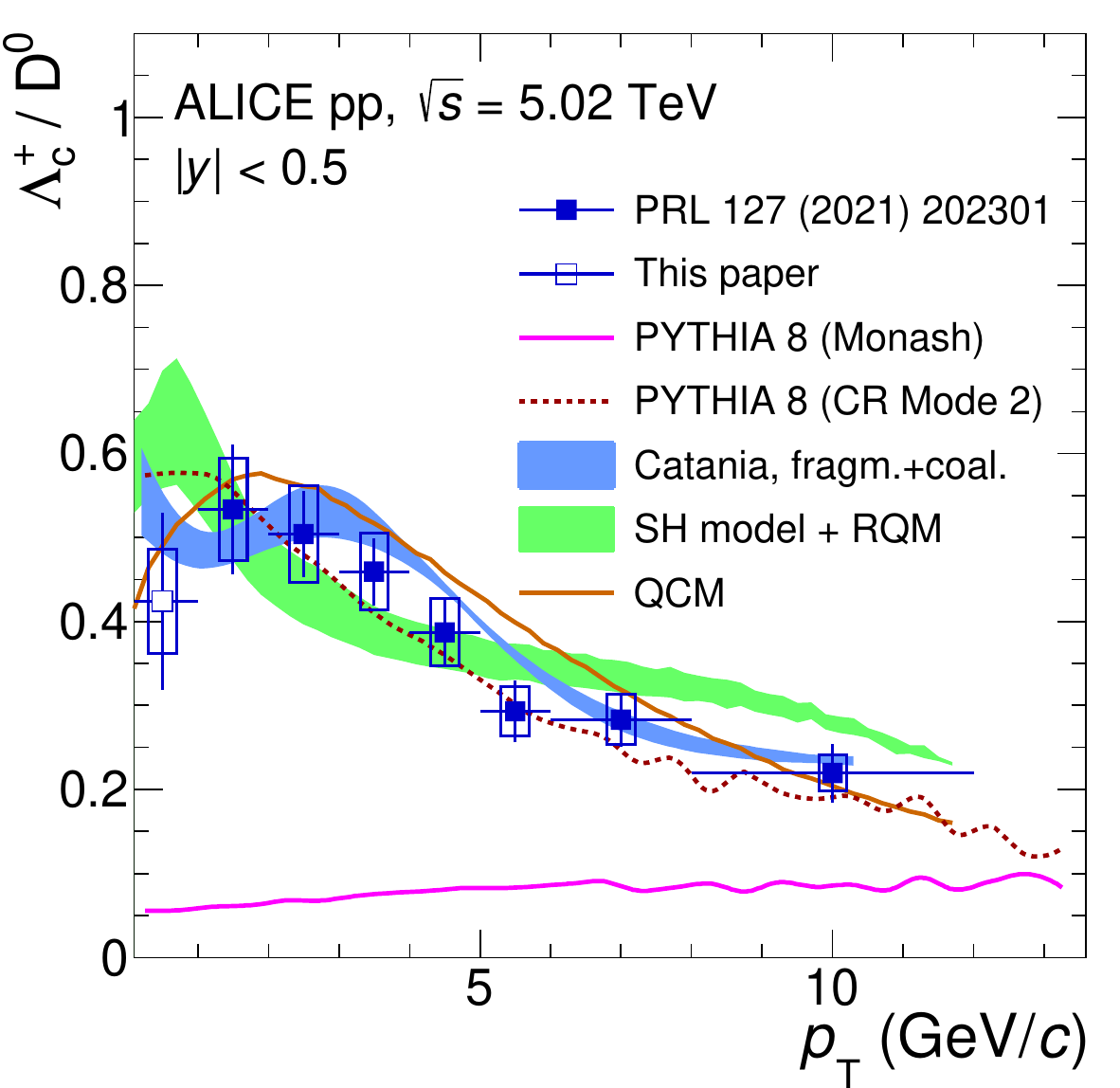}
\caption{Left panel: Charged $D$ meson production asymmetry $A$ as a function of Fermi momentum fraction $x_F$ as measured by the E791, WA82 and E769 
experiments. The experiments use charged pion beams 
with energies of 500, 340 and 250 GeV, respectively.
Reprinted from Ref.\ \cite{E791:1996htn} with permission from Elsevier.
Right panel: $\Lc/D^0$ ratio measured by ALICE in $pp$ collisions at $\sqrt{s}=5.02$ TeV compared with several model calculations using recombination \cite{Minissale:2020bif,He:2019tik,Song:2018tpv} or pure string fragmentation \cite{Skands:2014pea,Christiansen:2015yqa}.
Figure adapted from Ref.\ \cite{ALICE:2022exq} under 
a CC BY 4.0 license.} 
\label{fig:e791_CMS}
\end{figure}

While string fragmentation models enjoyed great success in the 1980s and 1990s, there were already hints in experimental data at the time that allowed for an interpretation in terms of quark recombination. Broadly speaking, these are measurements which show that in certain kinematic regions hadrons are preferentially produced with valence quarks which are very likely to have been present in the system prior to hadronization --- either from hard processes or beam remnants --- and not from string breaking or statistical processes. A good example is the leading-particle effect discussed in Sec.\ \ref{sec:2-1}. The left panel of Fig.\ \ref{fig:e791_CMS} shows results from the E791, WA82 and E769 experiments which used charged-pion beams on a fixed target and measured, among other things, open-charm mesons as a function of the momentum fraction $x_F$ in the direction of the pion beam. The charm quark in the $D$ meson is mostly produced in a hard process and is thus present prior to hadronization.
Taking a $\pi^-$ ($\bar u d$) beam as an example, one finds an asymmetry between $D^-$ ($\bar c d$) and $D^+$ ($c\bar d$) mesons, which increases to close to 1 as $x_F \to 1$, \ie, charm quarks are more likely to hadronize with beam quarks ($d$) rather than quarks that are not valence quarks in the beam particles. This is illustrated in the left panel of Fig.\ \ref{fig:e791_CMS} where the asymmetry parameter, $A=\frac{N_{D^-}-N_{D^+}}{N_{D^-}+N_{D^+}}$, is defined as the ratio (leading $-$ non-leading)/(leading $+$ non-leading) where the leading particles are defined as those with beam valence quarks. In the case of a $\pi^-$ beam the $D^-$ and $D^0$ ($c \bar u$) are leading particles and the
$D^+$ and $\bar D^0$ ($\bar c u$) are non-leading particles.


It should be noted that PYTHIA can describe the data after some tuning (see Ref.~\ref{fig:e791_CMS}), as strings do contain both quarks from hard processes and beam remnants in some configuration. However, it was pointed out that a recombination picture in which the heavy quark coalesces directly with beam partons is a more natural picture which can describe the data equally 
well~\cite{Braaten:2002yt,Rapp:2003wn}. 

\begin{figure}[t]
\includegraphics[width=0.53\linewidth]
{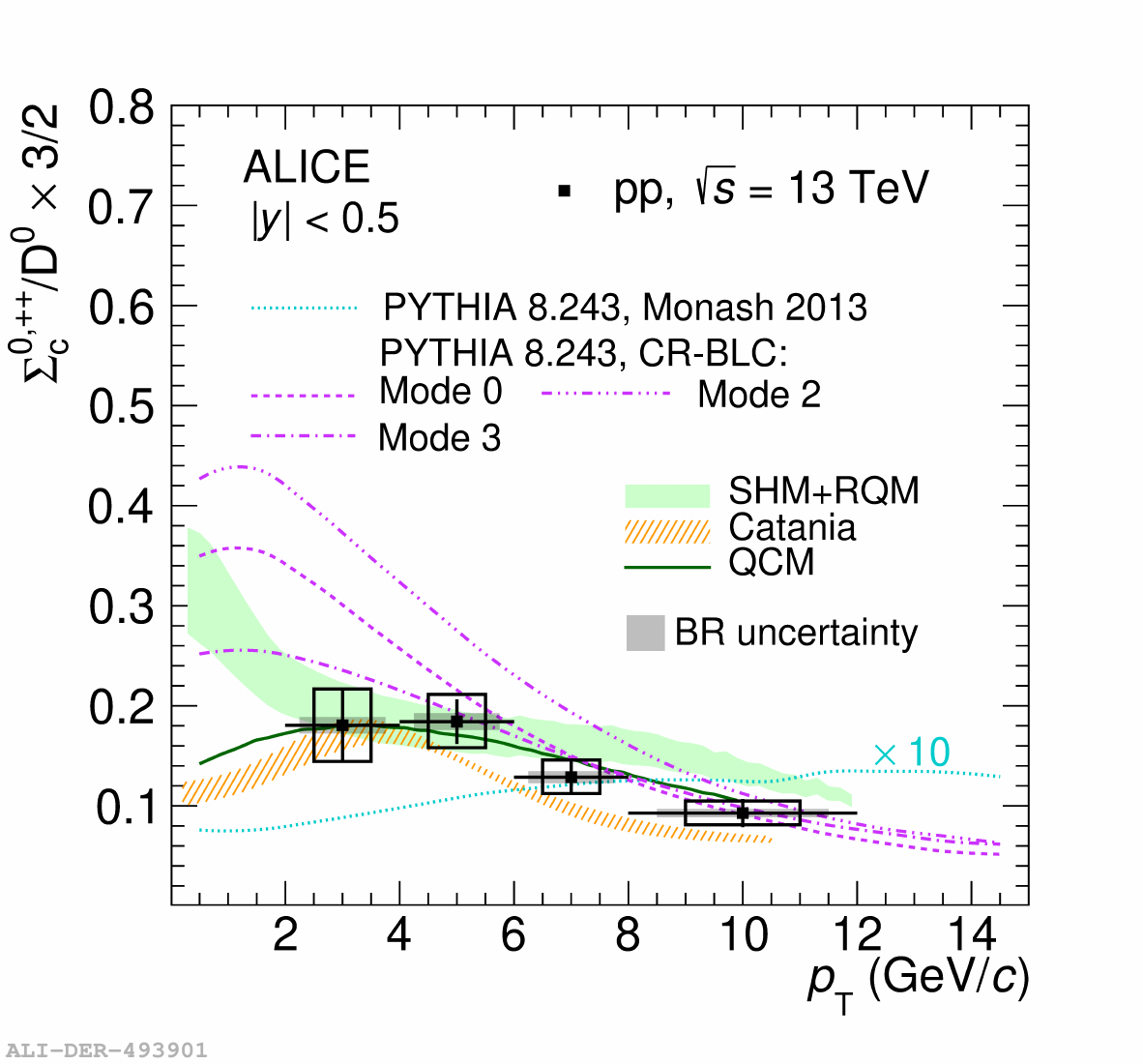}
\includegraphics[width=0.51\linewidth]
{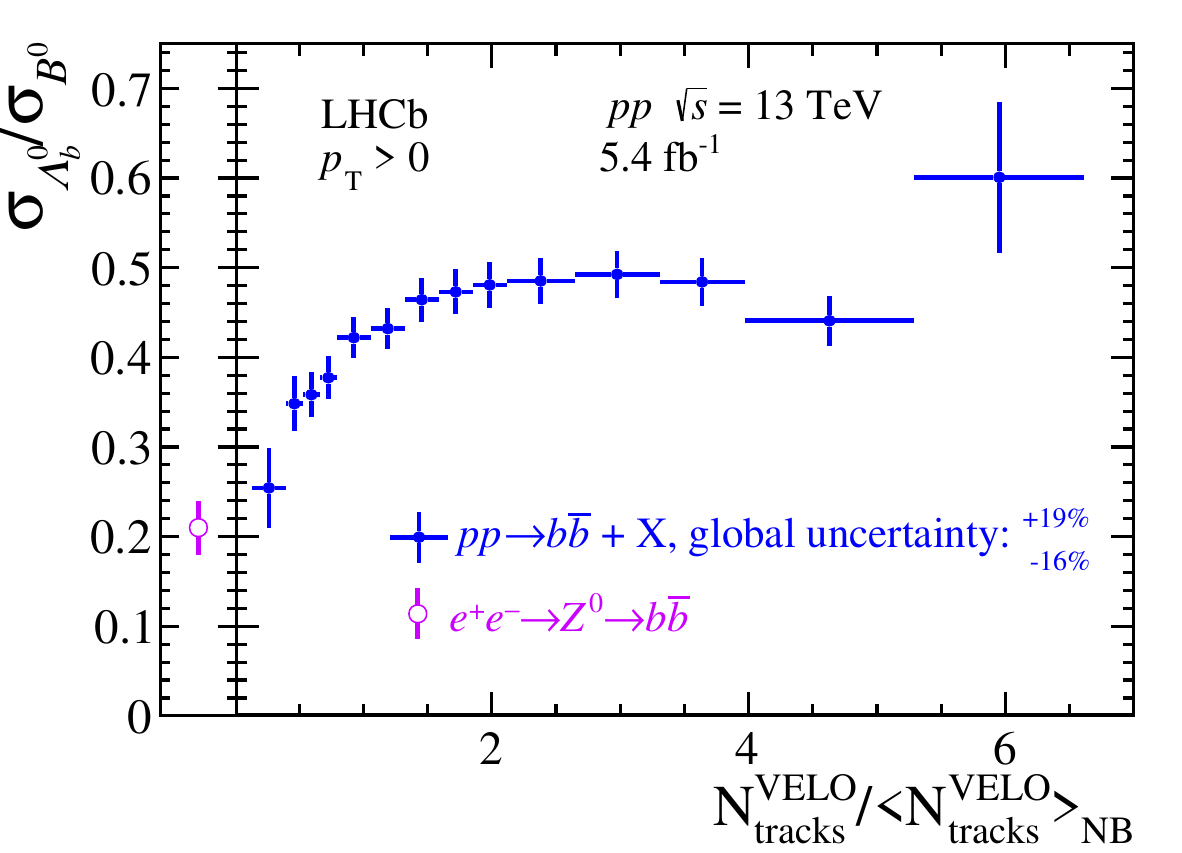}

\caption{Ratios of heavy flavor hadrons in $pp$ collisions at $\sqrt{s_{NN}}=13 \,\rm TeV$. Left Panel: $\Sigma_c^{0,++}/D^0$ vs $\pT$ at mid-rapidity compared to various models \cite{Skands:2014pea,Minissale:2020bif,He:2019tik,Song:2018tpv,Christiansen:2015yqa}. Figure reprinted from \cite{Altmann:2024kwx} under a CC BY 4.0 license. Right Panel: $\Lambda_b^0/B^0$ versus multiplicity of the events at forward rapidity compared to the value in $e^+e^{-}$ collisions \cite{LHCb:2023wbo}. Figure reprinted from Ref.\ \cite{LHCb:2023wbo} under a CC BY 4.0 license}
\label{fig:Sigmac_D0_pp}
\end{figure}

While recombination has become an accepted model in explaining certain features of data from heavy ion collisions over the past two decades, in recent years surprising additional hints for the existence of recombination mechanisms in elementary collisions like $pp$ have emerged. Measurements are showing  enhancements in baryon production which are not easily explained by string fragmentation. The most notable example comes from the open charm sector where the ratio of $\Lambda_c$ baryons to $D$ mesons is boosted over expectations from $e^+e^-$ collisions in a quite large range of momenta up to $p_T \simeq 10$ GeV/$c$, challenging the role of universality in fragmentation, see the discussion in Sec.\ \ref{sec:2-4}.
The right panel of Fig.\ \ref{fig:e791_CMS} shows a recent result from the ALICE experiment for $\Lc$ baryons to $D^0$ mesons which is not described well by standard PYTHIA 8 without adding additional color reconnections \cite{Skands:2014pea,Christiansen:2015yqa}. Several models based on quark recombination agree quite well with the data. As discussed in Sec.\ \ref{sec:2-4} they are also able to correctly predict the feed-down from $\Sigma^{0,++}$ resonances, as shown in the left panel of Fig.\ \ref{fig:Sigmac_D0_pp}. We note that PYTHIA with color reconnections predicts a very large ratio at intermediate and low $\pT$, an effect that can be traced back to the suppression of diquark states with $l=1$, as discussed at the end of Sec.\ \ref{sec:2-4}.

An important hint that parton multiplicity plays a role in hadronization was recently shown by the LHCb Collaboration. They observe that the large value of $\Lambda_b^0/B^0\simeq 0.5$ measured in $pp$ collisions even at forward rapidity recovers the value measured in $e^+e^-$ collisions for very low multiplicity events, while it saturates for intermediate and high multiplicity events \cite{LHCb:2023wbo}, as one would expect once recombination sets in, see right panel of Fig.\ \ref{fig:Sigmac_D0_pp}.

\subsection{Heavy-Ion Collisions}
\label{sec:3.2}

\subsubsection{Light Flavor}
\label{sec:3.2.1}

\begin{figure}[t]
    \centering
    \includegraphics[width=0.49\linewidth]{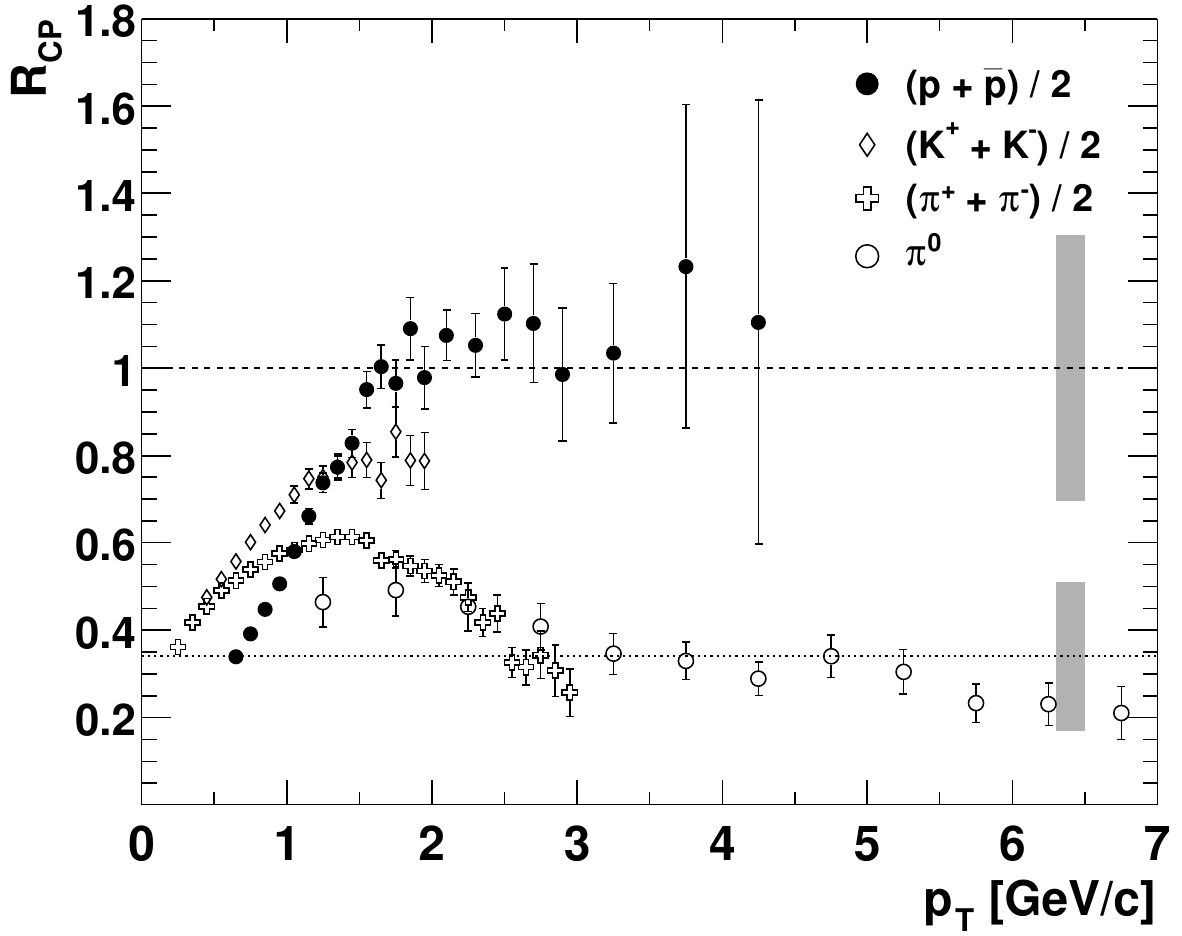}
    \includegraphics[width=0.49\linewidth]{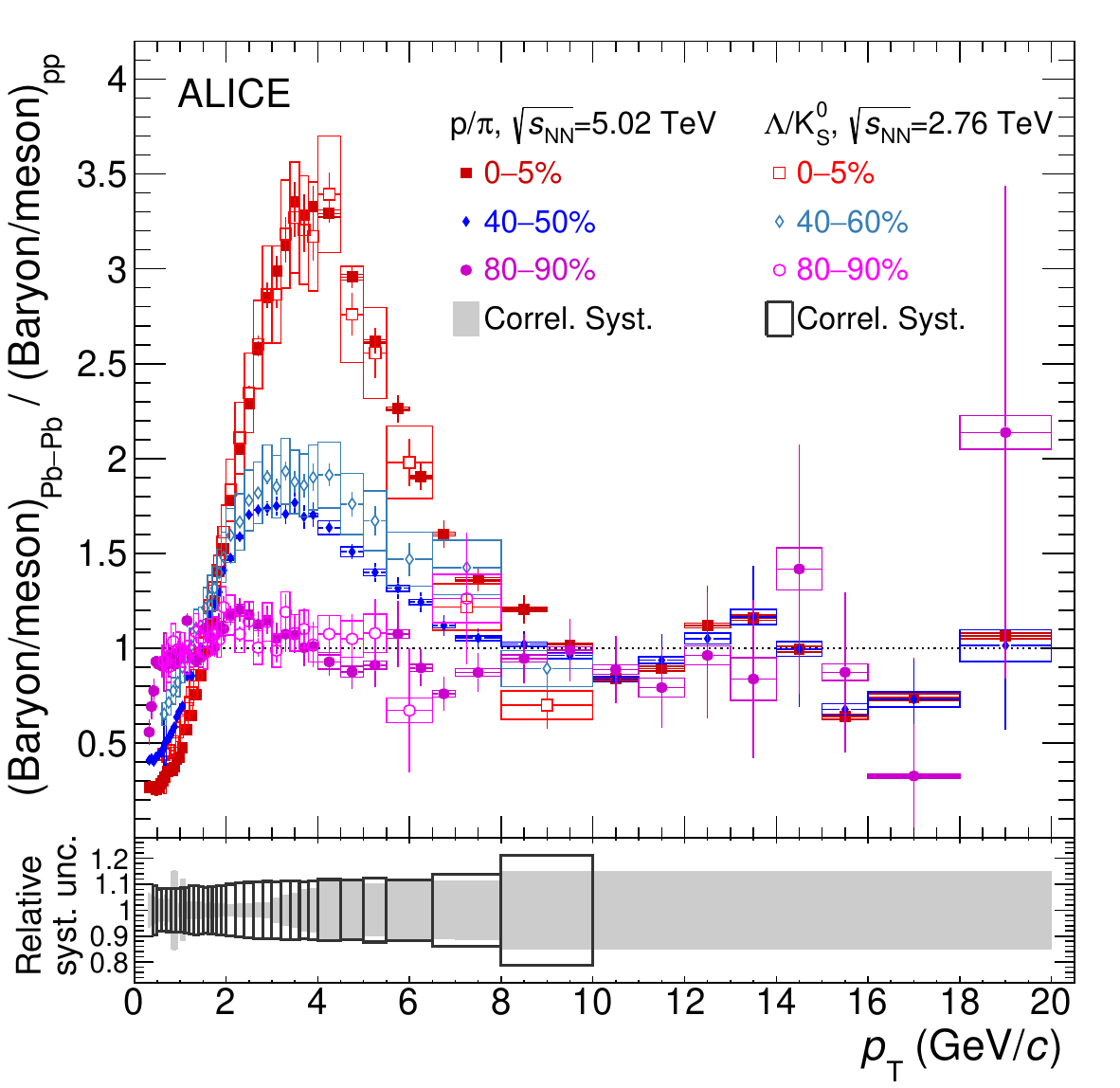}
\caption{Left panel: Early (2003) measurement of the PHENIX Collaboration of the central (0–10\%) to peripheral (60–92\%) ratios $R_\text{CP}$, scaled by the number of binary collisions, as a function of hadron $\pT$ for ($p$+$\bar p$)/2, charged kaons, charged pions, and neutral pions in Au-Au collisions at 200 GeV.
Reprinted from Ref.\ \cite{PHENIX:2003iij} with permission from the American Physical Society.
Right panel: Double ratios (baryon/meson ratios in Pb-Pb divided by the the same quantity in $pp$ at a same collision energy) measured as functions of hadron transverse momenta $\pT$ by the ALICE experiment.
Shown are $p/\pi$ at 5.02 TeV and $\Lambda/K_0^s$
at 2.76 TeV for several centrality bins.
Adapted from Ref.\ \cite{ALICE:2022wpn} under a CC BY 4.0 license.}
\label{fig:ALICE_PHENIX_RAA}
\end{figure}

The RHIC era began with several surprising features seen in early data. One was the apparent difference in the behavior of baryons and mesons observed in Au-Au collisions. In particular, the lack of suppression of baryons compared to mesons was deemed puzzling, as captured in a Science news article of the time \cite{Science:2002news}. One such early measurement by the PHENIX collaboration is shown in the left panel of Fig.\ \ref{fig:ALICE_PHENIX_RAA}. It displays the ratio $R_{\text{CP}}$ of the spectra of several identified hadrons in central to peripheral Au-Au collisions, scaled by the appropriate number of binary collisions. Pions and kaons show signs of jet quenching above 2 GeV/$c$ while protons stay close to 1 in the same region. We now understand that the onset of proton suppression is delayed to around 5-6 GeV/$c$ in central Au-Au collisions at RHIC energies, and jet quenching is rather similar for all light hadrons at very large $\pT$. 
The right panel of Fig.\ \ref{fig:ALICE_PHENIX_RAA} illustrates this universality between light hadrons. It shows the double ratio
\begin{equation}
    \frac{(B/M)_{\textrm{Pb-Pb}}}{(B/M)_{pp}}
    = \frac{R^B_{\textrm{Pb-Pb}}}{R^M_{\textrm{Pb-Pb}}}
\end{equation}
measured by ALICE for both $p/\pi$ and $\Lambda/K^0_s$ at 5.02 and 2.76 TeV, respectively for various centralities in Pb-Pb.
At these energies universality (identical suppression for all light hadrons) starts around 8 GeV/$c$ while baryons show an enhancement over mesons between 2 and 8 GeV/$c$ which is the same phenomenon noticed in the early RHIC data.

\begin{figure}[t]
    \centering
\includegraphics[width=0.75\linewidth]{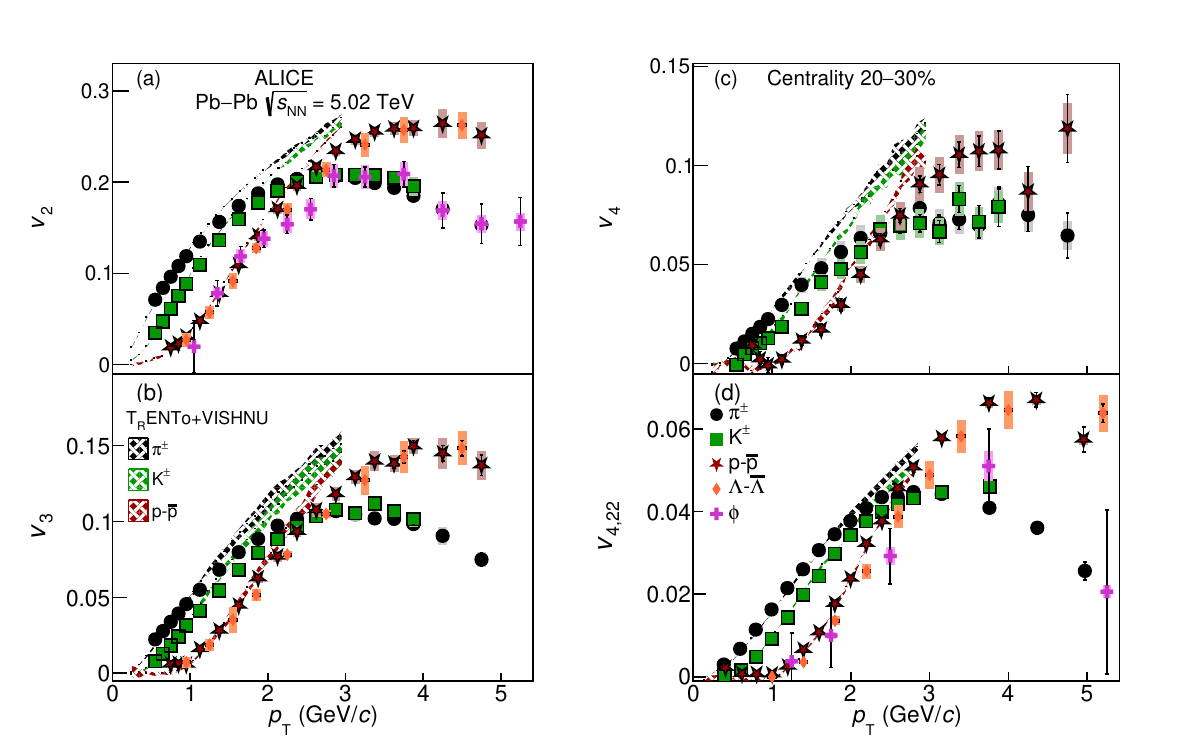}
\caption{The flow coefficients (a) $v_2$, (b) $v_3$, (c) $v_4$, and (d) $v_{4,22}$ as functions of hadron transverse momentum $\pT$ as measured by the ALICE experiment in the
20–30\% centrality bin in Pb-Pb collisions at 5.02 TeV. 
The curves represent estimations extracted from the VISHNU fluid dynamic code with TRENTO initial conditions, see Ref.\ \cite{ALICE:2022wpn}.
Adapted from Ref.\ \cite{ALICE:2022wpn} under a CC BY 4.0 license.}
\label{fig:ALICE_vn}
\end{figure}

The interesting task at the time was to explain why baryons behave differently at intermediate momenta of 2-6 GeV/$c$ at RHIC (and 2-8 GeV/$c$ at LHC energies). This led to a proliferation of quark recombination models from 2003 onwards \cite{Greco:2003xt,Fries:2003vb,Greco:2003mm,Fries:2003kq,Molnar:2003ff,Ravagli:2007xx,Lee:2007wr,Oh:2009zj,He:2010vw,ExHIC:2010gcb,Wang:2012cw,Cao:2013ita,Minissale:2015zwa,Plumari:2017ntm,Song:2018tpv,He:2019vgs,Cao:2019iqs} in which baryon production, just as in elementary systems, is aided by the presence of a background of quarks at hadronization, either in the underlying event (in collision systems like $pp$) or in the flowing thermal QGP medium (in AA). The key argument for the baryon enhancement at intermediate $\pT$ is that the boost from recombining partons is more effective. Consider partons at $\pt$ of 2 GeV/$c$ either in the flowing QGP or as part of a shower. They can recombine into 6 GeV/$c$ baryons or 4 GeV/$c$ mesons at roughly equal strength. On the other hand, a 6 GeV/$c$ meson would require two 3 GeV/$c$ partons which is much less likely with the prevailing steep momentum spectra, see the more detailed discussion in Sec.\ \ref{sec:2-2}. 

As an alternative, it was suggested that the boost for baryons does not come from the larger number of valence quarks, but from its larger mass via the flow that thermal partons acquire in AA collisions. In that case the velocity is similar for baryons and mesons in a given fluid cell, leading to larger baryon momenta. Indeed, models which combine fluid dynamic calculations with sufficiently large flow and fragmentation hadrons from jets could qualitatively reproduce key differences between protons and pions at intermediate $\pT$ \cite{Hirano:2003pw}; in this approach the baryon elliptic flow can be larger than the meson one due to a species dependence of the transition to independent fragmentation. Also, to what extent hydrodynamics, even with viscous corrections, remains a valid approach for hadron momenta of $\pT \sim 3-6$ GeV/$c$, is not obvious since deviations from the equilibrium distribution are estimated to be potentially quite large \cite{Plumari:2015sia,Luzum:2008cw}. A quantitative understanding of the transition from the hydrodynamic to the kinetic (or energy-loss) regime remains an open question where HF hadrons in particular are expected to provide key insights.

\begin{figure}[t]
    \centering
\includegraphics[width=0.75\linewidth]{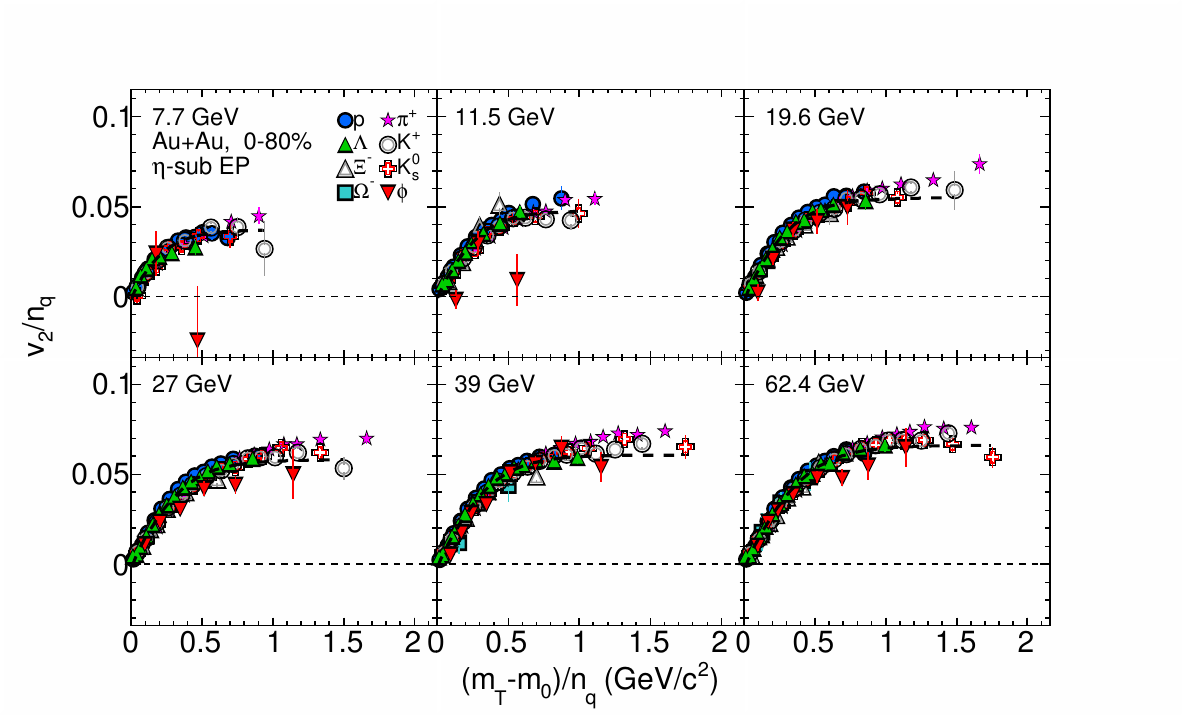}
\caption{
The elliptic flow scaled by the number of constituent quarks $v_2/n_q$ for several identified hadrons in Au-Au collisions in the 0–80\% central bin as a function of the constituent quark-scaled reduced transverse mass $(m_\text{T}-m_0)/n_q$ measured by the STAR experiment. 
Several collision energies (7.7 Gev -- 62.4 GeV) are shown together with a combined fit (excluding pions). The NCQ-scaled elliptic flow is consistent 
between particle species, but its fit differs from the same curve for their anti-particles ($\pi^-$, $\bar p$, etc; not shown here) at smaller collision energies. This suggests that within the group of particles/anti-particles constituent quark scaling holds to good accuracy. Note that the NCQ scaling of the $\phi$ meson is inconclusive below 19.6 GeV.
Reprinted from Ref.\ \cite{STAR:2013ayu} with permission from the American Physical Society.}

\label{fig:STAR_vn_BES}
\end{figure}

The two scenarios can be distinguished experimentally by comparing $\phi$ mesons and protons which have roughly the same mass but differ in their number of valence quarks. In a recombination picture $\phi$ mesons behave like other mesons, while in a pure flow picture they behave like protons. The data in the early 2000s was not precise enough for a final verdict. The situation has improved significantly since. Recent data sets, including elliptic flow data, paint a picture in which the $\phi$ meson unanimously behaves like a meson at both top RHIC and LHC energies, highlighting the role of the valence structure for hadronization. The most commonly cited experimental signature for this effect is the scaling of elliptic flow $v_2$ with the constituent quark number of the hadrons (NCQ-scaling). It was observed early in the RHIC program that the elliptic flow of identified hadrons follows the predictions of fluid dynamics for $\pT \lesssim 2$ GeV/$c$, which exhibit an ordering between hadron species according to mass. This is followed by a splitting between baryons and mesons at intermediate momenta above 2 GeV/$c$. Fig.\ \ref{fig:ALICE_vn}
shows several flow coefficients, including $v_2$ measured by ALICE, for five species of hadrons. Flow coefficients follow fluid dynamic predictions at low momenta which is followed by baryon-meson splitting at intermediate $\pT$. The $\phi$ meson clearly follows the behavior of kaons and pions, not protons.

If NCQ-scaled quantities are plotted, i.e.\ $v_2/n_q$ vs $\pT/n_q$, where $n_q$ is the number of valence quarks in a hadron, all particles fall more or less on a universal curve at intermediate momenta. This can be made even more pronounced by plotting data vs the scaled transverse mass $(m_\text{T}-m_0)/n_q$ which also collapses the points in the fluid dynamic regime on a universal curve. We demonstrate this feature in Fig.\ \ref{fig:STAR_vn_BES} for several species of hadrons (not including their anti-particles) for energies probed in the first RHIC energy scan. The scaling holds to good accuracy even at the smallest collision energies, albeit the behavior of the $\phi$ meson becomes less conclusive at the smallest energies.

NCQ-scaling is often referred to as a telltale signature for quark recombination. It persists through a very large range of collision energies and occupies the phase space between a purely thermal hadron phase and hadron production at very large momenta which is dominated by jets. It is then reasonable to assume, and it has been conjectured by many of the recombination models, that coalescence between quarks from jet showers and partons in the (nearly) thermalized QGP or underlying event, dominate the intermediate momentum range, extending, \eg, flow effects to surprisingly large momenta. On the theoretical side, the understanding of NCQ-scaling in recombination is straightforward under certain simplifying assumptions~\cite{Fries:2003kq}, and it appears to approximately persist when some of these assumptions are relaxed, see the discussion following Eq.\ \ref{eq:vn-scaling} in Sec.~\ref{sec:2-2}. However, NCQ scaling may not be considered as a general property of a recombination process because specific phase space distributions of quarks can lead to large violations~\cite{Pratt:2004zq}.

\subsubsection{Open Heavy Flavor}
\label{sec:3.2.2}

Open heavy-flavor particles are playing a key role in investigating hadronization processes in heavy-ion collisions. Quark recombination becomes a particularly clean process as the charm and bottom quark-masses are much larger than $\Th$ and $\Lqcd$ and thus conserve their identity (flavor quantum number) in the transition (since pair production is heavily suppressed). Early applications~\cite{Greco:2003vf} of coalescence models to $D$-mesons already indicated sizable effects in particular on their $v_2$. Subsequently, quantitative frameworks have been developed where heavy quark (HQ) transport simulations through the QGP generate realistic phase-space distributions that are implemented into subsequent hadronization models. 
\begin{figure}[t]
    \centering
 \includegraphics[width=0.47\linewidth, height=4.13 cm]{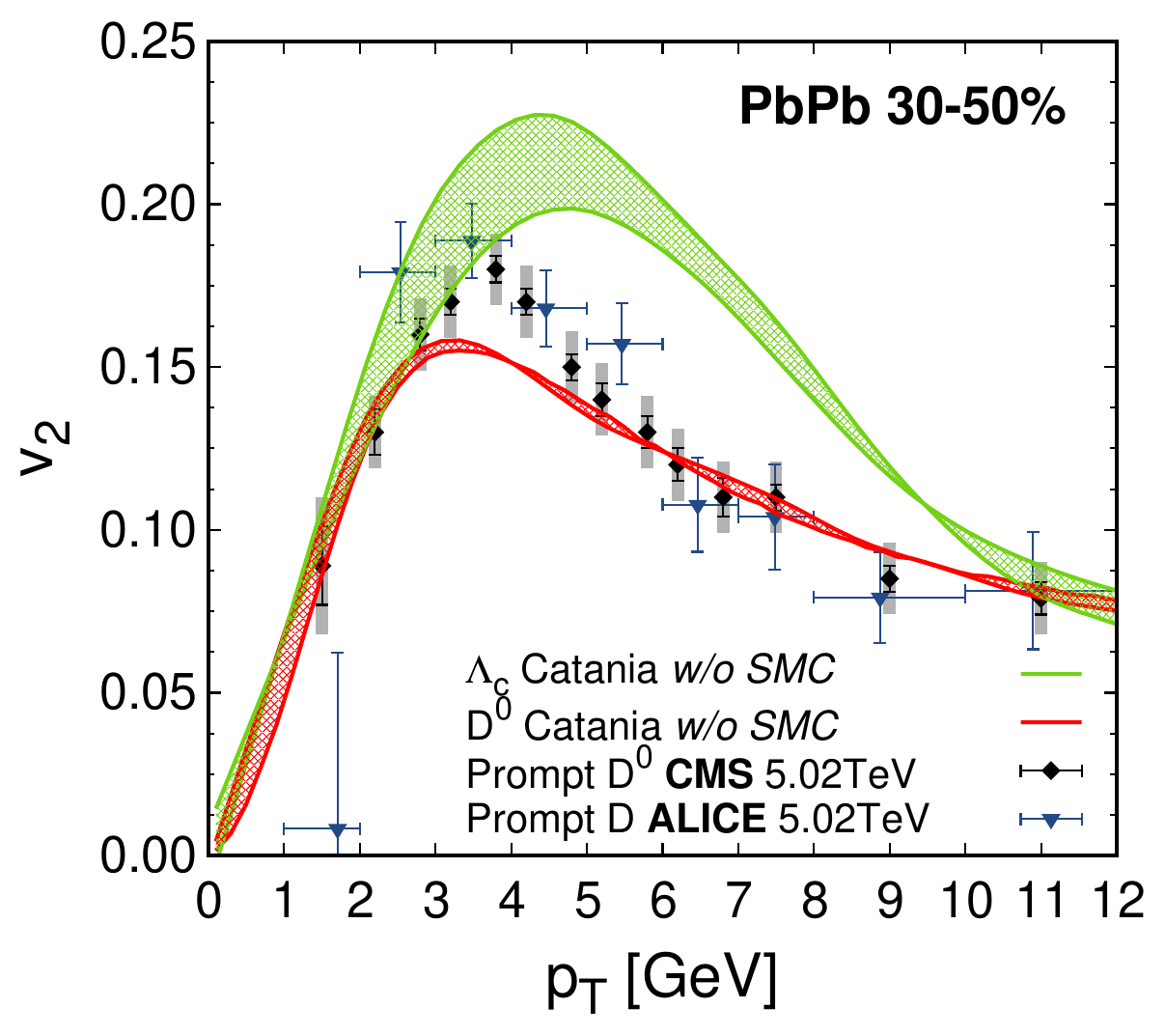}
 \includegraphics[width=0.523\linewidth, height=4.53cm]{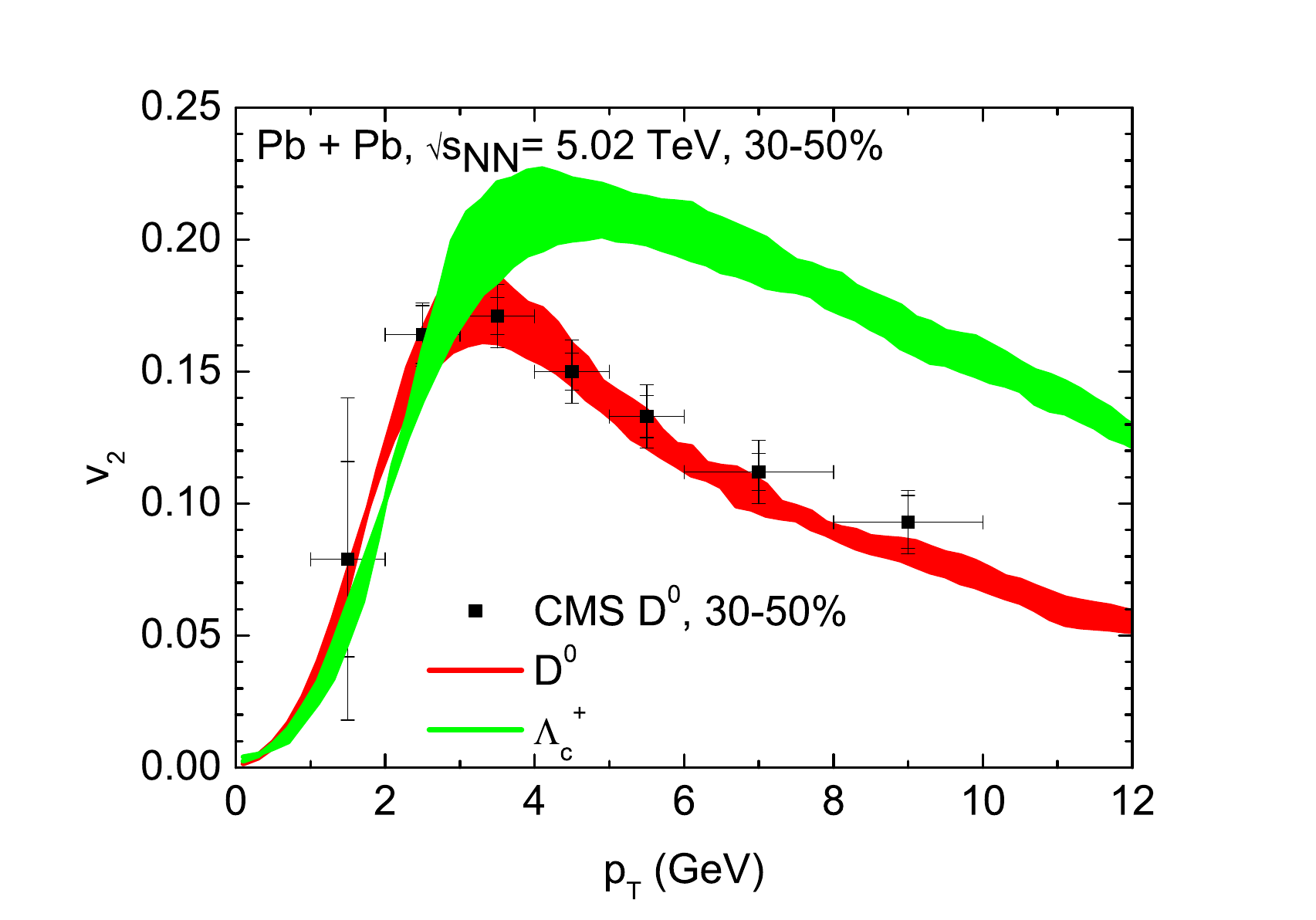}
\caption{
Elliptic flow of charm hadrons in Pb-Pb collisions at 5.02\,TeV (left and right panel) at the LHC. Calculations for $c$-quark transport in the QGP supplemented with phase-space coalescence from the Catania model w/o space-momentum correlation (SMC) for charm \cite{Plumari:2017ntm} 
 (left panel) and resonance recombination model including SMC~\cite{He:2019vgs} (right panel). These are compared to ALICE data ~\cite{Acharya:2018hre} 
 and CMS data, \cite{CMS:2020bnz} (left) and  ~\cite{Sirunyan:2017plt} (right), at $\sqrt{s_{NN}}=5.02$ for prompt D mesons. One recognizes the imprint of the hadronization mechanism on an underlying universal $c$-quark phase-space distribution function.
Adopted from Ref.~\cite{He:2019vgs} with permission from the American Physical Society.
}
\label{fig:v2-Lc-D}
\end{figure}
In this way, the conversion of the same underlying HQ distribution function leaves direct signatures in the yields and spectra of different types of HF hadrons. 
Subsequent developments have demonstrated that HQ recombination is a key mechanism in describing HF hadron observables in AA collisions at RHIC and the LHC. The first signature came from the challenge of a simultaneous description of the $v_2$ and the $\raa$ of HF observables in Au-Au collisions at RHIC \cite{Das:2015ana}, such as the semi-leptonic decay electrons at RHIC~\cite{PHENIX:2006iih}, which were successfully predicted by an early HQ transport/recombination approach~\cite{vanHees:2005wb}. This, in particular, also suggested the importance of non-perturbative interaction strength in the QGP which, over the years, has led to increasingly accurate extractions of the HQ diffusion coefficient of the QGP, a fundamental transport parameter of QCD matter.
Furthermore, it has also led to the realization that the same interactions that trigger the color-neutralization process through quark recombination~\cite{Ravagli:2007xx} may be at the microscopic origin of the non-perturbative interaction strength in the QGP in the form of dynamically emerging resonances in heavy-light scattering amplitudes~\cite{vanHees:2007me}. 
Recent LHC data on $D$-meson $\raa$ and $v_2$ have reached a precision that, through model comparisons, enabled an extraction of the HQ diffusion coefficient of
$1.5 < 2\pi\Ds \Tpc < 4.5$ in the vicinity of $\Tpc\simeq 160$\,MeV~\cite{ALICE:2021rxa}.
In particular, the HF hadron $v_2$ (see, \eg, Fig.~\ref{fig:v2-Lc-D}) is an excellent gauge of heavy-quark coupling strength to the medium as it is sensitive to the collectivity that develops starting from the initial distributions; thus a reliable decomposition of diffusion and hadronization effects is mandatory for a robust assessment of the transport coefficients.

The $v_2$ of $\Lc$ in AA collisions has not yet been measured, but it is likely accessible with Run 3 at the LHC. As can be seen from Fig.~\ref{fig:v2-Lc-D}, in a recombination picture it is expected to be larger than the $v_2$ of $D^0$ mesons at $\pT>\, 2-3$ GeV/$c$ for a wide range of $\pT$.
Comparing the left and right panels of Fig.\ \ref{fig:v2-Lc-D}, one notices that the extension of this $\pT$ range significantly depends on the inclusion of space-momentum correlations (SMCs) for the charm quarks. The SMCs for charm quarks manifest themselves at the freeze-out hypersurface essentially as a larger
presence of charm quarks at intermediate and high $\pt$ at larger $r$, where the bulk medium also carries larger radial and elliptic flow~\cite{He:2019vgs}.
If SMCs for charm quark are discarded, as in the modeling of Catania, see left panel of Fig.\ref{fig:v2-Lc-D}), one has $v_2(\Lc)\simeq v_2(D^0)$ at $\pT\simeq \, 10$ GeV/$c$, while the inclusion of SMCs, as in RRM-TAMU (right panel), the recombination process for $\Lc$ dominates over a larger range. In that case, at $\pT\simeq \, 10$ GeV/$c$ one still has significantly larger $v_2(\Lc)$ compared to $v_2(D^0)$.
Hence, a comparison of $v_2(\Lc)$ and the $v_2(D^0)$ is not only able to provide further tests of hadronization by coalescence, but also better insights into the charm-quark distribution in phase space. 

\begin{figure}[t]
    \centering
 \includegraphics[width=0.99\linewidth]{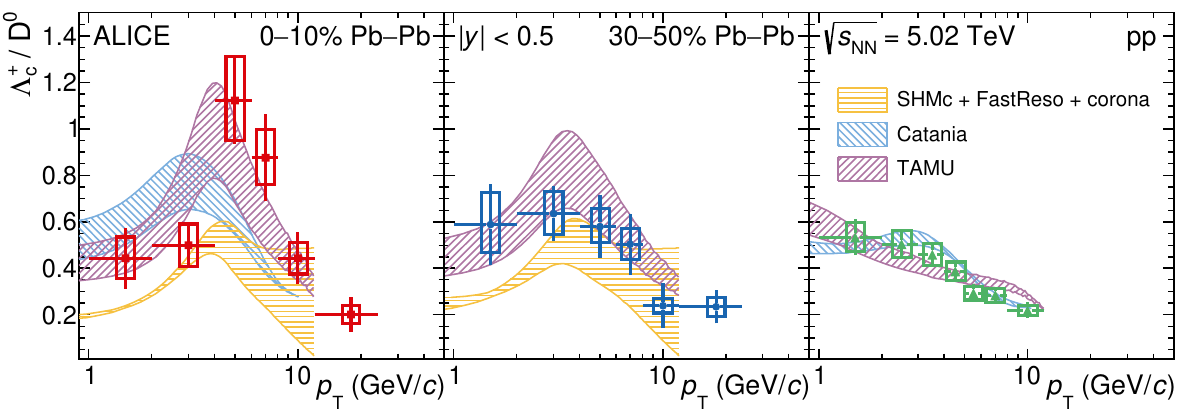}
\caption{Ratio of $\Lc$ over $D$-meson spectra in 5.02~TeV Pb-Pb (left and middle panels) and $pp$ collisions (right panel) at the LHC compared to Catania coalescence \cite{Minissale:2020bif}, TAMU recombination in AA \cite{He:2019vgs} and SHM-RQM in $pp$ \cite{He:2019tik} and statistical hadronization model (SHM) coupled to a fireball model \cite{Andronic:2021erx}.
Adopted from Ref.~\cite{ALICE:2022exq} under a CC BY 4.0 license. }
\label{fig:Lc-D}
\end{figure}
Measurements of the different types of charmed hadrons (hadro-chemistry) are becoming available with increasing precision also in AA collisions. For example, data for the production of $D_s=(c\bar{s})$ mesons in Au-Au~\cite{STAR:2021tte} and Pb-Pb collisions~\cite{ALICE:2021kfc} indicate an enhancement over the expectations from $pp$ collisions, which has been anticipated based on the coupling to a strangeness-equilibrated QGP (as opposed to a strangeness suppression in $pp$ collisions)~\cite{Andronic:2003zv,Kuznetsova:2006bh,He:2012df}. If the enhancement is indeed driven by coalescence, a rather prominent peak in the $\raa$ at relatively low $\pT$  has been predicted, which, however,  has not (yet) been observed in experiment.
It is of great interest, how the initially surprising charm-baryon production in $pp$ collisions at the LHC develops in a heavy-ion environment. It turns out that the integrated $\Lc/D^0$ ratio does not significantly change from $pp$ to semi-central and central Pb-Pb collisions. However, there are indications that a peak structure develops in central collisions~\cite{ALICE:2022exq}, see Fig.~\ref{fig:Lc-D} (similar observations have also been made in Au-Au collisions at RHIC~\cite{STAR:2019ank}). This is compatible with model calculations that have a substantial contribution to $\Lc$ production from recombination out to relatively large $\pT$ \cite{Plumari:2017ntm,He:2019vgs}. The peak structure is more prominent if aided by space-momentum correlations between fast-moving $c$-quarks and the outer layers of the hydrodynamically expanding medium where the light quark flow is large, as in the TAMU approach.  This effect further entails, as discussed above, a significantly larger $v_2$ of the $\Lc$ relative to the $D$-meson for $\pT\gsim$\,4\,GeV (see Fig.~\ref{fig:v2-Lc-D}), as two light quarks are required for recombination into a charm baryon. 
We also notice that the behavior of $\Lc/D^0$ is correlated to the difference between $v_2(\Lc)$ and $v_2(D^0)$ discussed above; in the Catania model, where SMCs have not been included, the $\pT$ range where $v_2(\Lc)>v_2(D^0)$ leads to a smaller peak structure of $\Lc/D^0>$ vs. $\pT$.
Future precision measurements of these quantities will test if a coalescence/recombination picture can correctly predict both $\Lc/D^0$ and $v_2(\Lc)/v_2(D^0)$ and also provide further extensions to other hadron species, and thus be invaluable to corroborate or refute these mechanisms.

\section{Quarkonia}
\label{sec:4}
\begin{figure}
    \centering
 \includegraphics[width=0.497\linewidth]{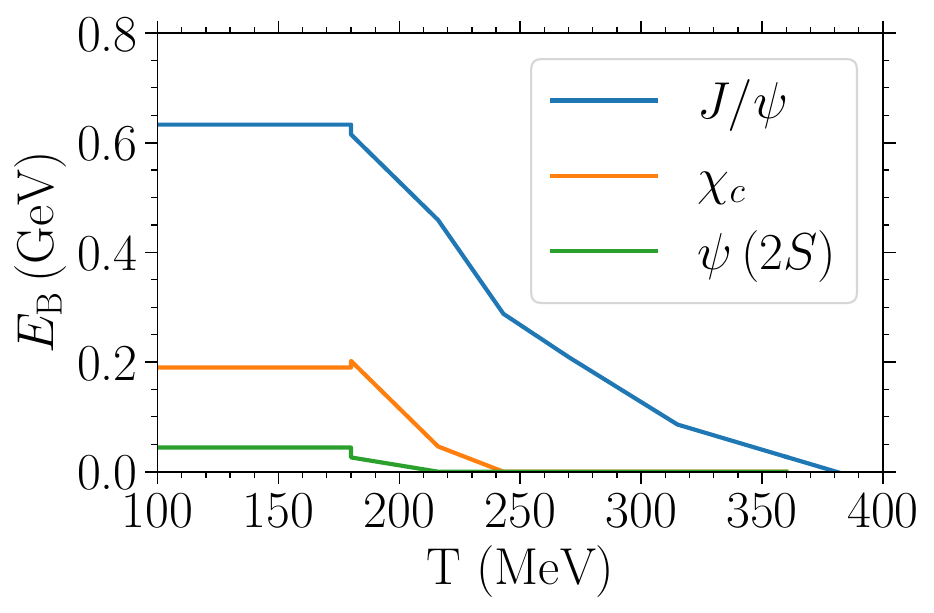}
 \includegraphics[width=0.497\linewidth]{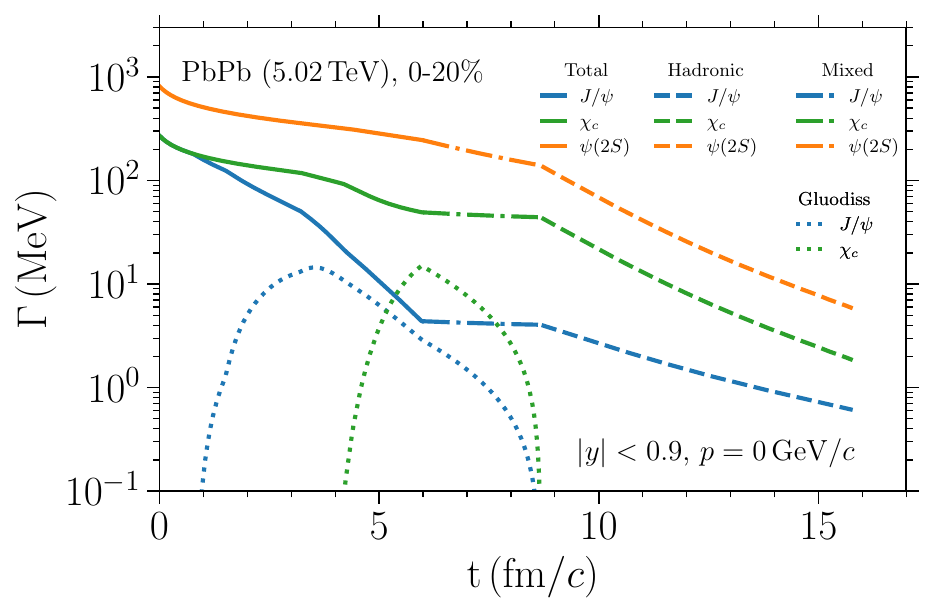}
\caption{Charmonium binding energies obtained from the internal-energy potential~\cite{Riek:2010fk} (left panel) implemented into rate calculations for the dissociation of $\jpsi$ (blue lines), $\chi_c$ (orange lines) and $\psi(2S)$ (green lines) mesons (right panel). The rates are plotted as a function of proper time in a central Pb-Pb collisions with an initial temperature of $\sim$550\,MeV, where the solid lines are the total rate in the QGP consisting of the (generally dominant) contribution from inelastic thermal-parton scattering and from gluo-dissociation (dotted lines); the dashed lines are from hadronic reactions (for $T<T_c$) while the dash-dotted lines follow from a mixed-phase construction of the QGP and hadronic rates at $T_c$=180\,MeV. Figures adopted from Ref.~\cite{Wu:2024gil} under a Creative Common CC BY license.}
\label{fig:Eb}
\end{figure}

Most of the coalescence approaches discussed above are based on, or implemented in, an instantaneous approximation. While the latter retains the important kinematic information on the preceding diffusion processes, it limits the dynamical information on the underlying interactions of the quarks and their coupling to the surrounding medium in the hadronization process. On the other hand, for quarkonia, significant efforts over the last 3 decades have been devoted to develop transport approaches that treat the time-dependent kinetics of their dissociation and (re-)generation, where reaction rates from inelastic scattering off the medium constituents play a key role~\cite{Thews:2000rj,Grandchamp:2003uw,Grandchamp:2005yw}. This was largely driven by the hypothesis that at least the more strongly bound quarkonia (like the ground-state $J/\psi$ and $\Upsilon(1S)$) can survive in the QGP up to certain temperatures loosely referred to as ``melting" temperatures below which ``regeneration reactions" set in~\cite{Grandchamp:2003uw,Grandchamp:2005yw}. Most of this work thus far has been carried out using semiclassical approaches based on the Boltzmann equation or kinetic rate equations of the type
\begin{equation}
N_{\cal Q} (t) =     \Gamma_{\cal Q} (N_{\cal Q} - \Neq) 
\end{equation}
The latter have the advantage that they manifestly encode the long-time limit in terms of the equilibrium abundance, $\Neq$ of each quarkonium state ${\cal Q}$. 
In practice, it is computed from their thermal density, $\Neq=V_{\rm FB} \gamma_{\cal Q} \gamma_{\bar Q} \int d^3k/(2\pi)^2 f^B(M_{\cal Q}; T)$, where $V_{\rm FB}$ is the time-dependent fireball volume (taken over a suitable range in rapidity, typically 1-2 units) and $f^B$ is the Bose distribution function of the quarkonium state. The $\gamma$ factors represent the HQ fugacities that are used to match the total HQ number in the system (dominated by the open HF states) to the number of produced $Q\Qbar$ pairs in the system. In a QGP the open HF degrees of freedom are essentially the charm or bottom quarks with their respective in-medium masses. In the hadronic phase one has to sum over all charmed hadrons which is in principle the same quantity used in the statistical hadronization model (see also the pertinent contribution in this volume).   
The inelastic reaction rates, $\Gamma_{\cal Q}$, represent the transport parameters for the dissociation and regeneration processes in the medium (for the latter also the equilibrium number is needed). 
Thus far the reaction rates have been mostly calculated via perturbative couplings to the medium. A key quantity in their evaluation is the binding energy, $E_B$, of each quarkonium state, which not only controls the magnitude of the rates (largely due to the available phase for the endothermic reactions which is smaller for larger binding energies) but also which type of processes gives the dominant contribution. Most contemporary calculations employ in-medium binding energies obtained from a nonperturbative input potential with screening~\cite{Andronic:2024oxz}. 
Parametrically, for large binding $E_B>T$,  2$\leftrightarrow$ 2 gluo-dissociation, $g+{\cal Q} \to Q +\Qbar$ is the leading process, while for $E_B\lesssim T$ the inelastic scattering processes, $p+{\cal Q} \to p Q +\Qbar$ (where $p$ denotes a anti-/quark or gluon from the heat bath) take over. 
In practice, however, there can be coefficients in these parametric dependencies which are (much) larger than one. For example, in the calculations for the $J/\psi$ discussed in Ref.~\cite{Wu:2024gil}, the rate from inelastic parton scattering takes over from gluo-dissociation at a temperature of $T\simeq$230\,MeV, where the in-medium $J/\psi$ binding energy is still $\sim$400\,MeV, see Fig.~\ref{fig:Eb}. 

\begin{figure}
    \centering
 \includegraphics[width=0.495\linewidth]{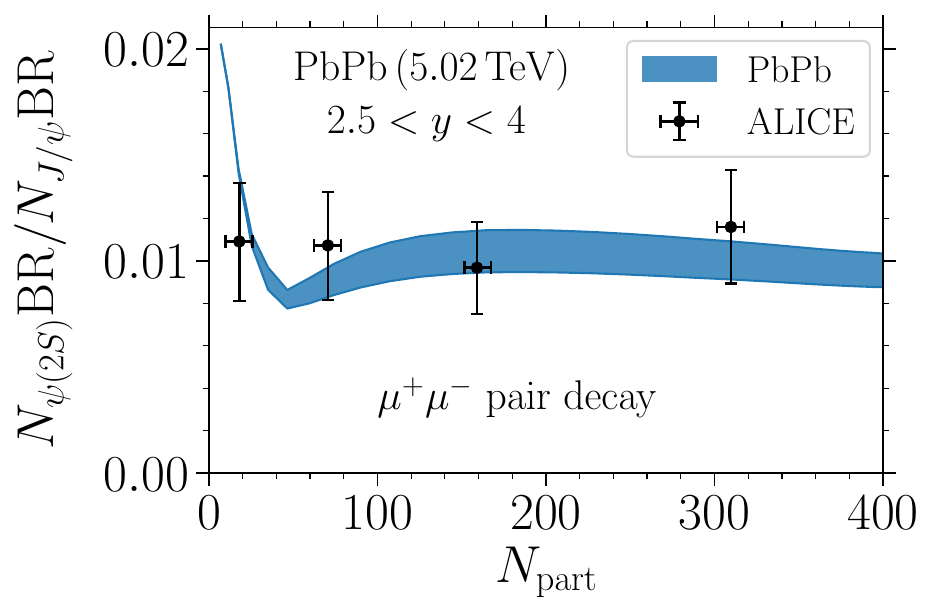}
 \includegraphics[width=0.495\linewidth]{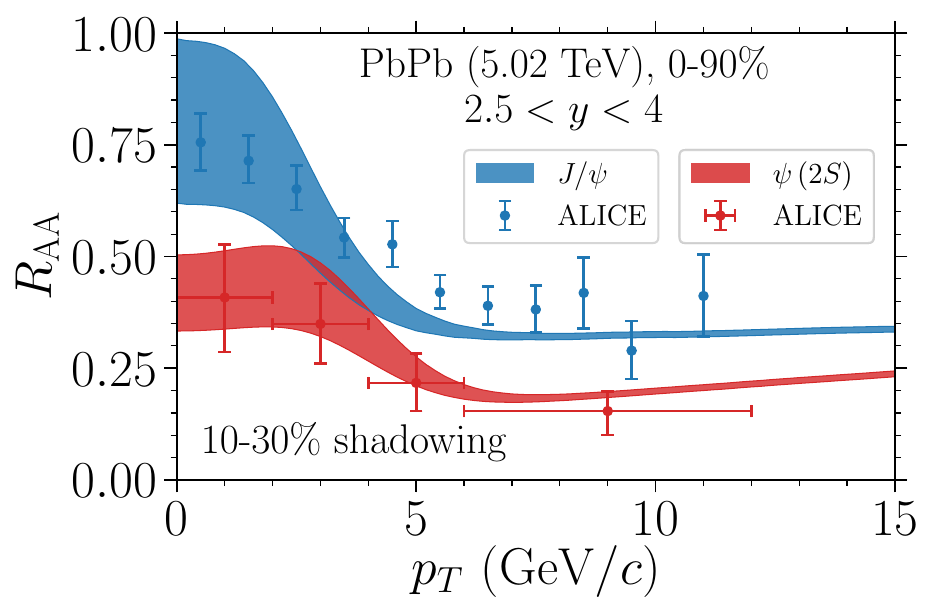}
\caption{$\jpsi$ and $\psip$ production in Pb-Pb(5.02TeV) collisions at the LHC at forward rapidity. Their ratio (including the branching fractions into the dimuon channel) as function of centrality is shown in the left panel, while right panel displays the individual nuclear modification factors as a function of $\pT$ for (near) minimum bias collisions. Data from the ALICE collaboration~\cite{ALICE:2022jeh} are compared to transport model calculations~\cite{Du:2015wha,Wu:2024gil} which incorporate a sequential-regeneration mechanism. Figures adopted from Ref.~\cite{Wu:2024gil} under a Creative Common CC BY license..
}
\label{fig:psi}
\end{figure}

Note that both processes are principally different from most of the coalescence models discussed in Sec.~\ref{sec:2}; the latter utilize 2$\to$ 1 hadron formation reactions which for on-shell particles require exothermic kinematics, \ie, $m_{q_1} + m_{q_2}> m_H$ (of course, in-medium spectral widths of the partons encoding the interactions with the medium can supply extra energy). Another key difference is that the quarkonium transport equations implement hadron formation as a continuous process, implying that there is a temperature hierarchy not only for the dissociation but also for the regeneration reactions; this led to the notion of ``sequential regeneration"~\cite{Du:2015wha} and was predicted to have observable consequences on the relation between $\jpsi$ and $\psip$ production. Specifically, the smaller binding of the $\psip$ entails that its dissociation temperature is much smaller than for the $\jpsi$, and its reaction rates are much larger especially at lower temperatures (even in the hadronic phase), cf.~Fig.~\ref{fig:Eb}. Thus, the $\psip$ is primarily produced later in the fireball evolution where the equilibrium limit tends to be larger than in earlier phases where most of the $\jpsi$ production occurs. In addition, later production implies that the recombining charm quarks have picked up additional collective flow from their diffusion in the medium which can create a maximum in the nuclear modification factor that is shifted to larger momentum for the $\psip$ relative to the $\jpsi$. These features turn out to be compatible with LHC data in Pb-Pb(5.02\,TeV) collisions, cf.~Fig.~\ref{fig:psi}. In particular, the observed $\psip/\jpsi$ ratio exceeds the equilibrium value of around $\sim$0.005 predicted by the statistical hadronization model~\cite{Andronic:2006ky,Andronic:2025jbp} while the results from sequential regeneration in a transport model calculation are closer to the data.

Similar mechanisms can be expected to be operative in bottomonium production in heavy-ion collisions, even though the magnitude of the regeneration contribution to the observed yields remains a more controversial issue than in the charmonium case at present. Compared to the charm case, $b\bbar$ production is much smaller, which for most centralities in Pb-Pb at 5.02\,TeV at the LHC leads to no more than one pair per unit rapidity (for charm more than 10 are expected in central collisions, which strongly enhances the probability for ``recombination"). However, the bottomonium production in $pp$ collisions is even more suppressed, with a ratio of $\Upsilon(1S)$ to $b\bar b$ of about 1-2\,$10^{-3}$, which is about a factor 10 smaller than for charmonium~\cite{Andronic:2015wma}. Thus, even a ``small" amount of regeneration of $\Upsilon$'s can have a very sizable effect on their $\raa$. A further complication is that the precise amount of regeneration depends on the momentum distribution of the heavy quarks and antiquarks in the system, \ie, their diffusion properties: a larger degree of kinetic thermalization entails  softer HQ spectra (in the thermal rest frame) which are more effective in regeneration~\cite{Grandchamp:2002wp,Song:2012at,Chen:2017duy,Yao:2020eqy,Du:2022uvj}. Since bottom quarks are $\sim$3 times heavier than charm quarks, their thermal relaxation time is about 3 times larger, typically 10\,fm/c~\cite{He:2022ywp}, which is comparable or even longer than the QGP lifetime in heavy-ion collisions. This provides both a challenge and an opportunity, since, on the one hand, it adds to the uncertainty in assessing the bottomonium regeneration yield while, on the other hand, it is directly connected to the open-bottom transport problem (including to bottom-quark hadronization discussed in the previous sections), thereby offering additional constraints.

Semi-classical approaches to quantum transport have achieved a rather comprehensive description of both charmonium and bottomonium production in heavy-ion collisions~\cite{Andronic:2025jbp}, predicting a dominant (sizable) contribution to charmonium (bottomonium) production from regeneration in Pb-Pb collisions at the LHC. In recent years, quantum transport frameworks have been developed, to provide a more accurate description of the regime where the quarkonium binding energy is small and thus effects of quantum formation times and energy uncertainty in both dissociation and regeneration can become relevant~\cite{Katz:2015qja,Yao:2018nmy,Akamatsu:2020ypb,Brambilla:2023hkw}, see \eg, the task force efforts summarized in Ref.~\cite{Andronic:2024oxz}. 
Further efforts are ongoing to implement the physics of the strongly QGP, in particular its strong coupling based on recent insights from lattice QCD~\cite{Tang:2023tkm} which suggest a very strong HQ potential with rather little screening in the QGP. The resulting non-perturbative quarkonium reaction rates are expected to be significantly larger than the ones implemented in current transport approaches, which likely will lead to a further increase in the recombination yields. Based on the same underlying interaction, the transport coefficients for the thermalization of individual heavy quarks are in the regime needed for the phenomenology of open HF production, in particular for the observed large values of the $D$-meson elliptic flow~\cite{Sirunyan:2017plt,ALICE:2021rxa}.

\begin{figure}[tbp]
\begin{center}
\includegraphics[width=0.7\linewidth]{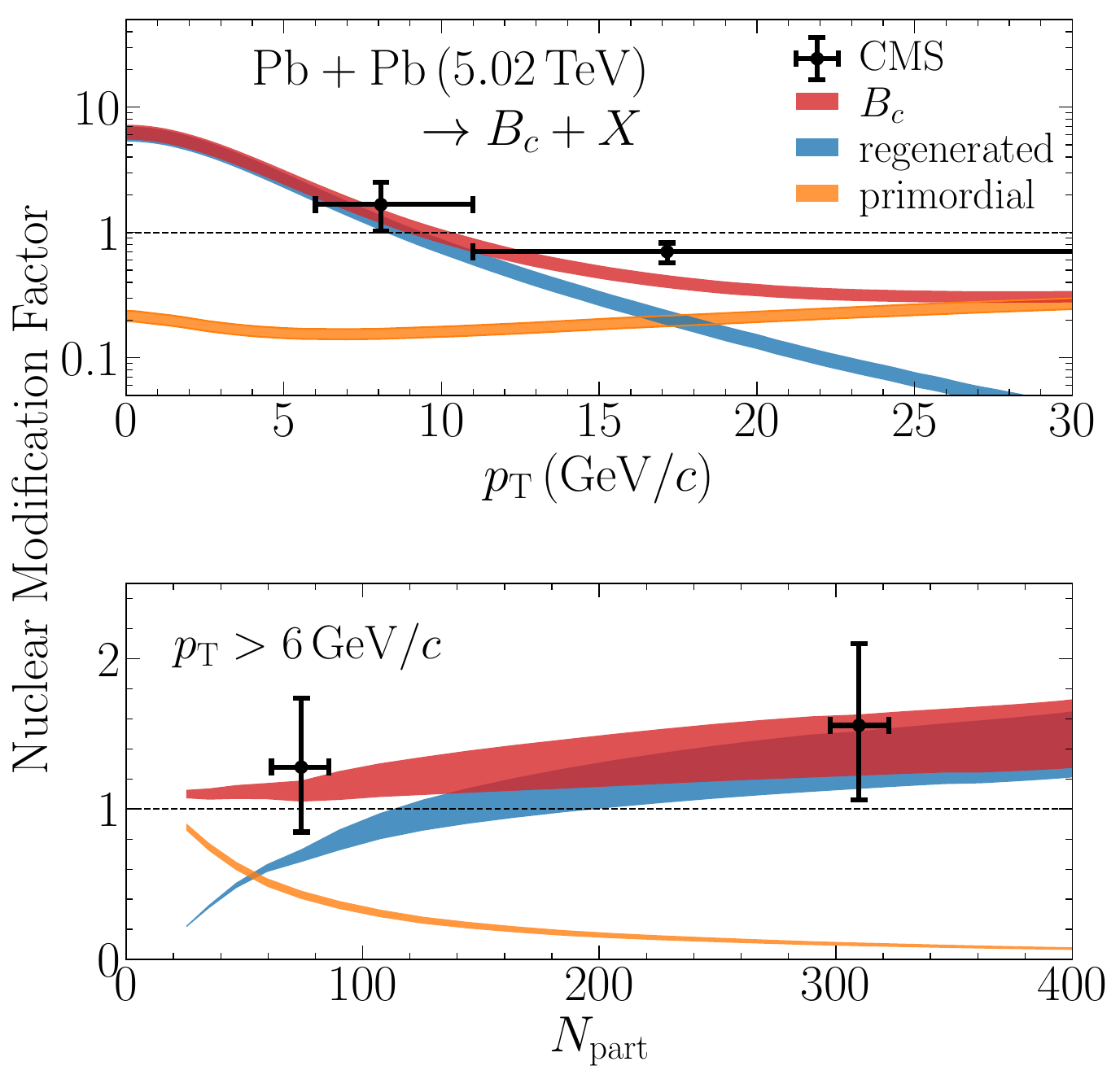}
\caption{Nuclear modification factor of $B_c$ mesons as a function of transverse momentum in Pb-Pb(5.02TeV) collisions at the LHC. Transport calculations~\cite{Wu:2023djn}, complemented with an coalescence model, are compared to CMS data~\cite{CMS:2022sxl}. One recognizes the large enhancement due to recombination processes at low $\pT$. The upper panel illustrates the uncertainty from $B_c$ production cross section in $pp$ collisions which is currently not well known; the lower panel indicates theoretical uncertainties in the radius parameters used in the coalescence model. Figure adopted from Ref.~\cite{Wu:2023djn} with permission from the American Physical Society.}
\label{fig:Bc}
\end{center}
\end{figure}
An even more sensitive probe of quark recombination are multiply heavy-flavored particles which are very difficult to produce in elementary reactions. For example, flavor conservation in the strong interaction does not allow for the direct production of $B_c$ mesons (which can only be done via $W$ decays in an electroweak process), and thus requires a double HF pair production followed by some sort of recombination. On the other hand, in a central Pb-Pb (5.02\,TeV) collision the production of a $b\bbar$ pair is accompanied by more than 10 $c\cbar$ pairs in same unit of rapidity, providing a rather favorable environment for $B_c$ formation. Early estimates of this mechanism~\cite{Schroedter:2000ek,Liu:2012tn} can nowadays be made much more quantitative based on the current understanding of both open and hidden HF transport in heavy-ion collisions. Utilizing a previously developed transport approach for charmonia and bottomonia~\cite{Wu:2024gil,Du:2017qkv} the kinetics of the ground and first-excited state for both $S$- and $P$-wave $B_c$ mesons have been calculated essentially without new parameters for Pb-Pb collisions at the LHC. The regeneration yield takes over from the suppressed initial production already in rather peripheral collisions (see lower panel in Fig.~\ref{fig:Bc} and leads to an inclusive (\ie, $\pT$ integrated) nuclear modification factor for the ground-sate $B_c(6280)$ of about 4-6 in central collisions. The $\pT$-dependence (summed over 0-90\% Pb-Pb centrality classes) has been obtained from a coalescence model with input $c$- and $b$-quark distributions from a state-of-the-art transport model; the results are shown in the upper panel of Fig.~\ref{fig:Bc} and illustrate the concentration of the enhancement at low $\pT\lsim m_{B_c}$; the sensitivity to the specific coalescence model is rather small (thanks to the large masses of the constituents). Clearly, the critical tests of theoretical predictions will measurements at low $\pT$.

\section{Summary}
\label{sec:5}
The formation of hadrons is a fundamental process
in nature that remains a challenging aspect of QCD to understand. Specifically, it is of pivotal importance to understand the conversion of the quark-gluon systems created in leptonic and hadronic collisions into colorless particles observable in experiment.
For elementary collisions, $e^+e^-$ and $e^-p$, and for particles at high momentum (jets), a successful description has been developed utilizing factorization theorems, in terms of an independent-fragmentation process where color neutral hadrons are formed through vacuum fluctuations of (anti-)quarks and diquarks. In the 1980s and 1990s, the independent-fragmentation picture has been complemented with string fragmentation models for particle production at low $\pT$. These mechanisms, based on "vacuum-like" hadronization starting from a single quark pair, are generally able to provide a robust description of particle production in elementary collision systems. 
However, particle production at forward and backward rapidity in both $pp$ and $\pi A$ collisions provided initial hints that quark hadronization can be affected by the presence of (valence) quarks in close-by phase space.

It was only in the early 2000s in AA collisions, where a very dense medium in phase space is created, that several signatures of coalescence of quarks from the medium as the dominant production mechanism at mid-rapidity have been observed. Specific features, mainly at intermediate $\pT\sim 2-6 \rm\,GeV$, were successfully described as a manifestation of quark coalescence into light hadrons. For open heavy-flavor production they were  found important as a dominant mechanism even at low $\pT$, especially also for charm baryons. 
A phase-space coalescence generally favors baryon production compared to "vacuum-like" mechanisms and generates a different pattern for mesons and baryons that manifests itself in a large baryon-to-meson ratio, separate branches of $R_\text{AA}(\pT)$ and/or $R_\text{CP}(\pT)$ and $v_2(\pT)$ between baryons and mesons. A fully microscopic and quantitatively satisfying description has not been fully formulated yet, although it appears to be in reach based on techniques developed during the last decade, such as a rigorous implementation of 4-momentum conservation or the development of realistic event-by-event transport simulations. Some aspects that have already been addressed should be further scrutinized,  such as the role of diquarks, the impact of space-momentum correlations, and the transition  to independent fragmentation in terms of the scales that determine the onset of the coalescence mechanism.
In this respect, the future heavy-ion program at the LHC will allow higher precision measurements down to very low $\pT$ along with new measurements especially for heavy-flavor hadrons, for example $\Xi_c$, $\Omega_c$, multi-charm baryons like  $\Xi_{cc}$ and $\Omega_{ccc}$ and the open bottom sector including $B_c$ mesons. When combined with system scans (both in system size and multiplicity) this will enable thorough insights into the dependence of hadronization mechanisms on the medium density and non-equilibrium aspects of incompletely thermalized heavy-quark distribution functions after their diffusion through a QGP of varying size and lifetime.

In the quarkonium sector, semi-classical transport models for the kinetics of $Q\bar Q$ mesons have been developed over the last two decades that have led to a rather robust description of available observables in both the charmonium and bottomonium sectors. Here, ``hadronization" is described as a continuous process whose onset can already occur rather deep into the QGP, depending on the in-medium binding properties of the various quarkonia. These approaches have already provided information on the underlying QCD force, with remnants of the confining force in the QGP playing an important role. However, these models have thus far largely focused on perturbative couplings to the medium, which is a shortcoming that is currently being addressed. In addition, open-quantum system approaches have become popular, necessary to enhance the reliability of the transport description in the regime where the in-medium binding energies are small. A challenge in these approaches remains the proper inclusion of regeneration mechanisms, and thus most of the applications have been carried out for bottomonia. To what extent the insights from the quarkonium transport models can be merged with mechanisms in the recombination models employed for light and open HF hadrons remains an open question.




%

%

%
\begin{acknowledgement}
RJF was supported by the U.S.\ National Science Foundation under grant numbers \rm{PHY-2111568}, \rm{PHY-2413003}, and \rm{OAC-2004571}.
The work of RR has been supported by the U.S.~National
Science Foundation under grant no. 
PHY-2209335 and by the U.S.
Department of Energy, Office of Science, Office of Nuclear
Physics through the Topical Collaboration in Nuclear
Theory on Heavy-Flavor Theory (HEFTY) for QCD
Matter under award no. DE-SC0023547. The work of VG has been supported by PRIN2022 under funding from  EU Next Generation (project code 2022SM5YAS). VG is grateful to C.M. Ko and P. Levai with whom he started the idea of phase-space quark coalescence and to V.\ Minissale and S.\ Plumari for all the work done together on this topic, and RR gratefully acknowledges the fruitful collaborations in particular with Min He.
\end{acknowledgement}
\end{document}